\documentclass[twocolumn,showpacs,preprintnumbers,amsmath,amssymb]{revtex4}

\usepackage{graphicx}
\usepackage{dcolumn}
\usepackage{bm}
\usepackage{epsf}

\renewcommand{\bar}[1]{\overline{#1}}
\newcommand{\ie}{{\it i.e.}}

\usepackage{amssymb}

\usepackage{indentfirst}
\usepackage{psfig,color}
\usepackage{epsfig}
\usepackage{epsf}
\usepackage{graphicx}

\providecommand{\Journal}[4] {#1 {\bf #2}, #3 (#4)}
\providecommand{\NPA}{Nucl. Phys. A } %
\providecommand{\NPB}{Nucl. Phys. B } %
\providecommand{\PLB}{Phys. Lett. B } %
\providecommand{\PRD}{Phys. Rev. D } %
\providecommand{\PRSA}{Proc. Roy. Soc. A } %
\begin{document}

\title{Exotic baryons with charm number $\pm1$ from Skyrme model }

\author{Bin Wu}
\affiliation{Department of Physics, Peking University, Beijing
100871, China}
\author{Bo-Qiang Ma}
\altaffiliation{Corresponding author}\email{mabq@phy.pku.edu.cn}
\affiliation{ CCAST (World Laboratory), P.O.~Box 8730, Beijing 100080, China\\
Department of Physics, Peking University, Beijing 100871, China}

\begin{abstract}
We illustrate the exotic $SU(3)$ baryon sub-multiplets in the
$SU(4)$ baryon multiplets predicted from the flavor $SU(4)$
collective-coordinate quantization, and investigate the exotic
states with charm number $C=\pm1$ up to leading order of $1/N_c$
under chiral $SU(3)_L\times SU(3)_R$ symmetry and the heavy quark
limit from bound state approach. We find that there exist the
$\bar{15}$-plets and 24-plets with $C=1$ and the $\bar{6}$-plets,
$15$-plets and $15^{\prime}$-plets with $C=-1$ bounded in this
approach, qualitatively consistent with
 predictions from the flavor $SU(4)$ collective-coordinate
quantization. By fitting one unique parameter of leading $SU(3)$
flavor symmetry breaking term with the mass difference between
$\Xi_c$ and $\Sigma_c$ and up to $O(m_Q^0N_c^{-1})$, we give all
the average masses of baryons in $\bar{6}$-plet and $15$-plet.
Several general relations of these masses without any parameter
are also introduced.
\end{abstract}

\pacs{12.39.Mk; 12.39.Dc; 12.40.Yx; 14.20.Lq }

\vfill

\vfill

\maketitle
\par
\section{INTRODUCTION}

Skyrme's soliton picture of nucleons \cite{Skyr} has been
successfully extended to the case of three light flavors of quarks
by directly generalizing collective-coordinate method from $SU(2)$
to the $SU(3)$ case \cite{Guad}. And there is an alternative
approach that hyperons are treated as bound states of solitons and
$K$ mesons by Callan and Klebanov (CK) \cite{Ca_Kl}. CK method was
first suggested to be applicable to charmed or bottomed baryons in
\cite{rrs}. And Walliser \cite{Wall} extended the
collective-coordinate method in the flavor $SU(3)$ Skyrme model to
the general $SU(N_f)$ case. However, because the mass of the heavy
flavor quarks $m_{c,b,t}$ is larger than
$\Lambda_{QCD}$($\sim200~MeV$) in contrast with that of the light
flavors $m_{u,d,s}$, the physical world respects the approximate
light flavor $SU(3)$ chiral symmetry more reliably. A physical
system involving both light flavors($u,d,s$) and heavy
flavors($c,b$) has the heavy quark flavor symmetry and the heavy
quark spin symmetry in the heavy quark limit
($m_Q\rightarrow\infty$), as well as the chiral symmetry for the
light quark system, \ie, the dynamics of the system is unchanged
under the exchange of heavy quark flavors and under arbitrary
transformations on the spin of the heavy quarks. In this limit,
the spin of heavy quark $\mathbf{S}_Q$ is conserved as well as the
total angular momentum $\mathbf{J}$, thus the spin of light
degrees of freedom $\mathbf{S}_l=\mathbf{J}-\mathbf{S}_Q$ is also
conserved. Combined with these symmetries, the CK approach has
been successfully extended to describe the static properties of
the heavy baryons with one charm or bottom quark
\cite{JAM,GLM,GMSS,OPM}, and also been used to chiral $SU(3)$ case
\cite{mss}.
\begin{figure}
\begin{center}
\includegraphics[width=8cm]{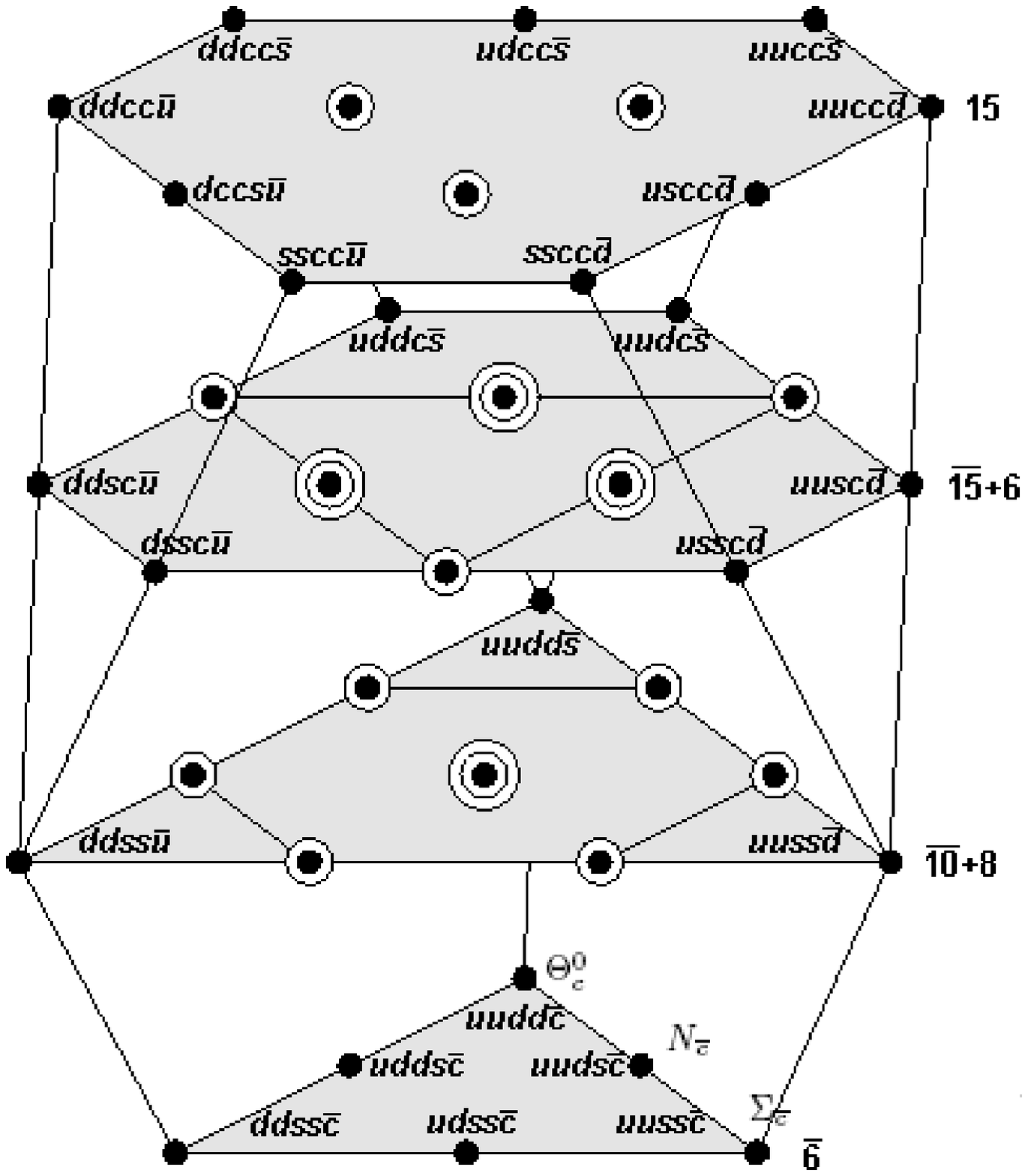}
\end{center}
\caption{The exotic pentaquark states in
 the $SU(4)$ $\bar{60}$-plet}\label{fig1}
\end{figure}
\par
Both the flavor $SU(4)$ collective-coordinate quantization and the
bound state approaches predicted the existence of charmed baryon
multiplets with exotic baryons. From the $SU(N_f)$ symmetric
collective Hamiltonian \cite{Wall}, we can see that the
$\bar{60}$-plet with spin $S$=1/2 and the 140-plets with
$S$=(3/2,1/2) are the lightest $SU(4)$ baryon multiplets next to
the $SU(4)$ 20-plet with spin 1/2 and 20$^\prime$-plet with spin
3/2 \cite{RGG}. We illustrate the weights of the $SU(4)$
representations $\bar{60}=(0,2,1)$ and $140=(2,1,1)$ in
Figs.~\ref{fig1} and \ref{fig2}, where the minimal quark contents
of the exotic states are also suggested. In the $SU(4)$
$\bar{60}$-plet, there exists a light flavor $SU(3)$ baryon
multiplet $\bar{15}$-plet with $C=1$ and $S=1/2$ and a
$\bar{6}$-plet with $C=-1$ and $S=1/2$, while in the $SU(4)$
140-plets, there are $\bar{15}$-plets with $C=1$ and
$S=(3/2,1/2)$, 24-plets with $C=1$ and $S=(3/2,1/2)$ and
$15$-plets with $C=-1$ and $S=(3/2,1/2)$. These $SU(3)$ baryon
multiplets all contain exotic states with $C=\pm1$. Bound state
approaches also predicted the existence of heavy antiquark states
with minimal five-quark configurations $uudd\bar{c}$ and
$uudd\bar{b}$ in the quark language~\cite{RS,OPM1}.
\par
The main purpose of this paper is to study the exotic states with
charm number $C=\pm1$ from bound state approach in three light
flavor case under chiral $SU(3)_L\times SU(3)_R$ symmetry and the
heavy quark limit. There are two approaches to describe heavy
baryons as bound states of solitons and heavy mesons. In
\cite{GLM,JAM}, solitons are first quantized to produce light
baryon states. Then, heavy baryons are constructed as bound states
of these light baryons and heavy mesons with explicit spin and
isospin. This is different from the approach in \cite{GMSS,OPM},
where heavy-meson-soliton bound states are first found and then
quantized by the collective coordinate method to give baryon
states with heavy quarks. In this paper, we investigate exotic
states with one anti-quark ($\bar{c}$) as well as those with one
heavy quark ($c$) and compare the results with those in
Figs.~\ref{fig1} and \ref{fig2}, predicted from the flavor $SU(4)$
collective-coordinate quantization, and give the mass spectra of
the $\bar{6}$-plet and 15-plet.
\par The paper is organized as follows. In Sec.~II, we give both the
Lagrangian of a physical system only involved with heavy mesons
containing one flavor of heavy quark and that with their
anti-particles. Then, in Sec.~III, we briefly review the approach
in \cite{JAM} under chiral $SU(3)_L\times SU(3)_R$ symmetry and
the heavy quark limit, and generalize it to describe baryons with
one heavy anti-quark ($\bar{c}$). After that, we study the exotic
states with $C=\pm 1$ by this approach, and calculate the average
masses of the $\Theta^0_c(uudd\bar{c})$ and
$\Theta^*_c(uddd\bar{c},uudd\bar{c},uuud\bar{c})$ states, which
belong to the $\bar{6}$-plets with $S_l=1$ and 15-plet with
$S_l=1$ in Figs.~\ref{fig1} and \ref{fig2} respectively. In
Sec.~V, we quantize the chiral $SU(3)$ symmetric effective
Lagrangian and give the mass spectra of the $\bar{6}$-plet and
15-plet baryons. And in Sec.~VI, we compare the results with those
in Figs.~\ref{fig1} and \ref{fig2} and give our conclusion.
\newpage
\begin{widetext}
\begin{center}
\begin{figure}
\includegraphics[width=15cm]{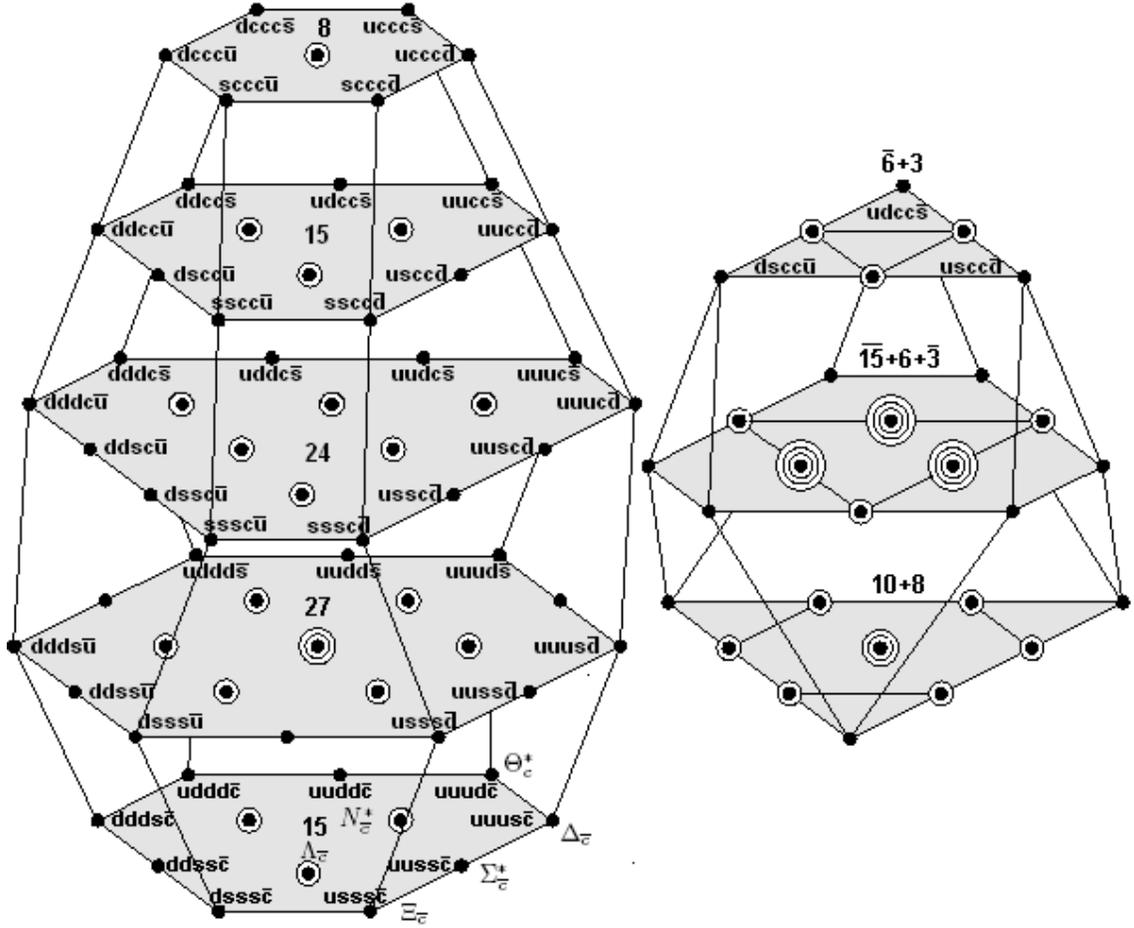}
\caption{The exotic pentaquark states in the $SU(4)$
140-plet}\label{fig2}
\end{figure}
\end{center}
\end{widetext}
\section{The EFFECTIVE LAGRANGIAN}
We first introduce the effective Lagrangian which respects both
chiral symmetry of the light flavors and heavy flavor and spin
symmetry. The $SU(3)$ symmetric action for the light flavor system
is of the form
\begin{equation}\begin{array}{rl}
I_l=&\frac{f^2_\pi}{4}\int d^4x\mbox{Tr}(\partial_\mu
\Sigma\partial_\mu \Sigma^\dagger)\\ &+\frac{1}{32e^2}\int
d^4x\mbox{Tr}\left[\partial_\mu \Sigma\Sigma^\dagger,\partial_\nu
\Sigma\Sigma^\dagger\right]^2+N_c\Gamma\equiv\int
d^4x\mathcal{L}_l,\label{Ll}
\end{array}
\end{equation}
where $f_\pi\approx 93$~MeV is the observed pion decay constant;
the second term is the so-called Skyrme quartic term, which
contains the dimensionless parameter $e$ to stabilize the
solitons; $\Gamma$ is the Wess-Zumino-Witten term; the chiral
fields $\Sigma(x)$ are elements of $SU(3)$.
\par The heavy meson field for the ground state $Q\bar{q}_a$ ($0^-$
and $1^-$) mesons can be represented by a $4\times4$ matrix field
$H_a$ \cite{wise}
\begin{equation}
H_a=\frac{1+v\llap{$/$}}{2}[P^*_{a\mu}\gamma^\mu-P_a\gamma_5],
\end{equation}
where $P_a$ and $P_{a\mu}^*$ represent the pseudoscalar ($0^-$)
and the vector ($1^-$) meson fields respectively, which annihilate
the $s_l=1/2$ meson multiplet and move with a four velocity
$v_\mu$. However, $P_a$ and $P_{a\mu}^*$ do not create the
corresponding antiparticles by the particle projector
$\frac{1+v\llap{$/$}}{2}$ in the heavy quark limit, thus, a new
field $\widetilde{H}_a$ is introduced to describe the
antiparticles of the heavy mesons $q_a\bar{Q}$ \cite{OPM1}
\begin{equation}
\widetilde{H}_a=[\widetilde{P}^*_{a\mu}\gamma^\mu-\widetilde{P}_a\gamma_5]\frac{1-
v\llap{$/$}}{2},
\end{equation}
where $\widetilde{P}_a$ and $\widetilde{P}^*_{a\mu}$ annihilate
the antiparticles of the $s_l=1/2$ meson multiplet and move with a
four velocity $v_\mu$. The fields $H_a$ and $\widetilde{H}_a$
transform as a (1/2,1/2) representation of $S_Q\otimes S_l$ and
$S_l\otimes S_Q$ respectively. Under $SU(2)_Q\otimes SU(2)_l$,
$H_a(x)$ and $\widetilde{H}_a$ transform as
\begin{eqnarray}
H_a&\rightarrow& S_QH_aS_l^\dagger,\label{ql}\\
\widetilde{H}_a&\rightarrow&
S_l\widetilde{H}_aS_Q^\dagger,\label{lq}
\end{eqnarray}
where $S_Q\in SU(2)_Q$ and $S_l\in SU(2)_l$. Similarly, under the
isospin transformation, $I_H\in SU(2)_I$, we have
\begin{eqnarray}
H_a&\rightarrow& H_bI_{Hba}^\dagger,\label{hi}\\
\widetilde{H}_a&\rightarrow& I_{Hab}\widetilde{H}_b,\label{ih}
\end{eqnarray}
\par For considering the couplings of heavy mesons to the
pseudo-Goldstone bosons, it is convenient to introduce
\begin{equation}
\xi=\sqrt{\Sigma},
\end{equation}
which under the $SU(3)_L\times SU(3)_R$ chiral symmetry transforms
as
\begin{equation}
\xi\rightarrow L\xi U^\dagger=U\xi R^\dagger,
\end{equation}
where $U$ is a function of $L$, $R$ and the chiral field. Then,
under chiral symmetry, $H_a$ and $\widetilde{H}_a$ transform as
\begin{eqnarray}
H_a&\rightarrow& H_bU^\dagger_{ba},\\
\widetilde{H}_a&\rightarrow& U_{ab}\widetilde{H}_b.
\end{eqnarray}
By including allowed terms with one derivation for the
heavy-meson-light-meson interaction, the Lagrangians which respect
both chiral symmetry and heavy quark spin and flavor symmetry are
of the form~\cite{wise,OPM1}
\begin{eqnarray}
\mathcal{L}=\mathcal{L}_l-iv_\mu\mbox{Tr}(D^\mu
H\overline{H})+g\mbox{Tr}(H\gamma^\mu\gamma_5A_\mu\bar{H}),\\
\widetilde{\mathcal{L}}=\mathcal{L}_l-iv_\mu\mbox{Tr}(\overline{\widetilde{H}}D^\mu
\widetilde{H})+g\mbox{Tr}
(\bar{\widetilde{H}}\gamma^\mu\gamma_5A_\mu\widetilde{H}),
\end{eqnarray}
with
\begin{eqnarray}
D_\mu H&=&H(\overleftarrow{\partial}_\mu+V_\mu^\dagger)=H(\overleftarrow{\partial}_\mu-V_\mu),\\
D_
\mu\widetilde{H}&=&(\partial_\mu+V_\mu)\widetilde{H},\\
\overline{H}^a&=&\gamma^0H_a^\dagger\gamma^0,\bar{\widetilde{H}}^a=\gamma^0\widetilde{H}_a^\dagger\gamma^0,
\end{eqnarray}
where $V_\mu$ and $A_\mu$ are defined as
\begin{eqnarray}
V_\mu&\equiv&\frac{1}{2}(\xi^\dagger\partial_\mu\xi+\xi\partial_\mu\xi^\dagger),\label{vmu}\\
A_\mu&\equiv&\frac{i}{2}(\xi^\dagger\partial_\mu\xi-\xi\partial_\mu\xi^\dagger)\label{amu}.
\end{eqnarray}\par
However, because of the singularity in $\xi$, it is convenient to
redefine $H_a$ as~\cite{JAM}
\begin{equation}
H_a^\prime=H_b\xi_{ba},
\end{equation}
and, similarly, we redefine $\widetilde{H}_b$ as
\begin{equation}
\widetilde{H}_a^\prime=\xi_{ab}^\dagger\widetilde{H}_b.
\end{equation}
Then, the Lagrangians become
\begin{eqnarray}\nonumber
\mathcal{L}=\mathcal{L}_l-iv_\mu\mbox{Tr}(\partial^\mu
H^\prime\overline{H}^\prime)+\frac{i}{2}v_\mu\mbox{Tr}(H^\prime\Sigma^\dagger\partial^\mu
\Sigma\bar{H}^\prime)\\
+\frac{ig}{2}\mbox{Tr}(H^\prime\gamma^\mu\gamma_5\Sigma^\dagger\partial_\mu\Sigma\bar{H}^\prime),\label{L}\\
\nonumber
\mathcal{\widetilde{L}}=\mathcal{L}_l-iv_\mu\mbox{Tr}(\overline{\widetilde{H}}^\prime\partial^\mu
\widetilde{H}^\prime)-\frac{i}{2}v_\mu\mbox{Tr}
(\bar{\widetilde{H}}^\prime\Sigma^\dagger\partial^\mu
\Sigma\widetilde{H}^\prime)\\
+\frac{ig}{2}\mbox{Tr}(\bar{\widetilde{H}}^\prime\gamma^\mu\gamma_5\Sigma^\dagger\partial_\mu\Sigma\widetilde{H}^\prime)\label{Lt}.
\end{eqnarray}
Then, under the transformations (\ref{ql})-(\ref{ih}), we have the
following operators from above Lagrangians
\begin{eqnarray}
S^k_{lH}&=&\int
d^3x\mbox{Tr}(\frac{1}{2}H^\prime\sigma^k\bar{H}^\prime),\label{slf}\\
S^k_{l\widetilde{H}}&=&-\int
d^3x\mbox{Tr}(\frac{1}{2}\bar{\widetilde{H}}^\prime\sigma^k\widetilde{H}^\prime),\\
I^k_H&=&\int
d^3x\mbox{Tr}(\frac{1}{2}H^\prime\lambda^k\bar{H}^\prime),\\
I^k_{\widetilde{H}}&=&-\int
d^3x\mbox{Tr}(\frac{1}{2}\bar{\widetilde{H}}^\prime\lambda^k\widetilde{H}^\prime),\label{ihtf}
\end{eqnarray}
where
$\sigma^i\equiv\frac{1}{4}i\epsilon^{ijk}[\gamma^j,\gamma^k]$,
$i,j,k=1,2,3$, and $\lambda^k$ are the tree generators of the
$SU(3)$ group.

\section{The Formulism for heavy baryons from bound state approach}

Under the assumption of maximal symmetry, there exists a solitonic
solution (with unit baryonic charge) of the equation of motion
from (\ref{Ll}) \cite{Guad}
\begin{equation}
    \Sigma_1(\mathbf{x})=\left(
        \begin{array}{cc}
            \exp{[i(\mathbf{\widehat{r}}\cdot\mathbf{\tau})F(r)]} & \begin{array}{c}0\\0\end{array}\label{heghog}\\
            \begin{array}{cc}0&0\end{array}&1
        \end{array}
        \right),
\end{equation}
where $F(r)$ is the spherical-symmetric profile of the soliton
with $F(0)=\pi$ and $F(\infty)=0$, $\mathbf{\tau}$ are the three
Pauli matrices, and $\mathbf{\widehat{r}}$ is the unit vector in
space. The action is invariant under the $SU(3)_L\times SU(3)_R$
transformation. However, in order to describe soliton solutions
with the same vacuum, it is only necessary to deal with those with
the form:
\begin{equation}
    \Sigma(x)=A(t)\Sigma_1(\mathbf{x})A(t)^{-1},\ \  A \in SU(3)\label{cllect}.
\end{equation}
With this assumption, after quantizing Eq.~(\ref{Ll}) in
collective coordinate $A$, we will get the $SU(3)$ symmetric
Hamiltonian \cite{Guad} and from it we can see the lowest
representations of light solitons (baryons) are the octet with
$s_l$=1/2, the decuplet with $s_l$=3/2, the anti-decuplet with
$s_l$=1/2, the 27-plet with $s_l$=3/2,1/2, etc. The light baryon
wave functions in the collective coordinates $A$ are
\begin{equation}
    \Psi_{\nu\nu^\prime}^{(\mu)}(A)=\sqrt{\mbox{dim}(\mu)}D^{(\mu)}_{\nu\nu^\prime}(A)\label{psi},
\end{equation}
where $(\mu)$ denotes an irreducible representation of the SU(3)
group; $\nu$ and $\nu^{\prime}$ denote $(Y, I, I_3)$ and $(1, J,
-J_3)$ quantum numbers collectively; $Y$ is the hypercharge of the
corresponding light baryons; $I$ and $I_3$ are the isospin and its
third component respectively; $J_3$ is the third component of spin
$J$; and $D^{(\mu)}_{\nu\nu^\prime}(A)$ are representation
matrices of the $SU(3)$ group.
\par Besides the collective
coordinate quantized chiral symmetric Hamiltonian, the interaction
Hamiltonian of Eq.~(\ref{L}) by semiclassical approximation is
\cite{GLM}
\begin{eqnarray}\nonumber
H_I^\prime=\frac{gF^\prime(0)}{2}\int d^3x\mbox{Tr}(
H^\prime\gamma^i\gamma_5A\lambda_iA^\dagger\bar{H}^\prime)\\
=2gF^\prime(0)I^k_HS^j_{lH}D^{(8)}_{kj},\label{hip}
\end{eqnarray}
and from Eq.~(\ref{Lt}), we have
\begin{eqnarray}\nonumber
\widetilde{H}_I^\prime=\frac{gF^\prime(0)}{2}\int d^3x\mbox{Tr}(
\bar{\widetilde{H}}^\prime\gamma^i\gamma_5A\lambda_iA^\dagger \widetilde{H}^\prime)\\
=-2gF^\prime(0)I^k_{\widetilde{H}}S^j_{l\widetilde{H}}D^{(8)}_{kj},\label{whip}
\end{eqnarray}
with \cite{GLM}
\begin{eqnarray}
I^k_HH&=&-H\frac{\tau^k}{2},\\
S^k_{lH}H&=&-H\frac{\sigma^k}{2},
\end{eqnarray}
and
\begin{eqnarray}
I^k_{\widetilde{H}}\widetilde{H}&=&\frac{\tau^k}{2}\widetilde{H},\\
S^k_{l\widetilde{H}}\widetilde{H}&=&\frac{\sigma^k}{2}\widetilde{H},
\end{eqnarray}
where $D^{(8)}_{kj}$ is the adjoint representation of the SU(3)
group and defined as
\begin{equation}
D^{(8)}_{kj}(A
)=\frac{1}{2}\mbox{Tr}(A^{\dagger}\lambda_kA\lambda_j),
\end{equation}
$S^k_{lH}$, $I^k_H$, $S^k_{l\widetilde{H}}$ and
$I^k_{\widetilde{H}}$ are the quantum mechanics operators
corresponding to (\ref{slf})-(\ref{ihtf}). We can see that the
interaction $\widetilde{H}_I^\prime$ is the negative of
$H_I^\prime$, this is because
\begin{eqnarray}
H^\prime\gamma^i\gamma_5&=&-H^\prime\sigma^i,\\
\gamma^i\gamma_5\widetilde{H}^\prime&=&\sigma^i\widetilde{H}^\prime,
\end{eqnarray}
and we define
\begin{equation}
H_I=2V_II^k_HS^j_{lH}D^{(8)}_{kj},
\end{equation}
with
\begin{equation}
V_I=\left\{\begin{array}{cc}gF^\prime(0),&H_I=H_I^\prime;
\\-gF^\prime(0),&H_I=\widetilde{H}_I^\prime.\end{array}\right.
\end{equation}
\par It is convenient to define
\begin{equation}
\mathbf{K}\equiv \mathbf{I}_H+\mathbf{S}_{lH}.
\end{equation}


The allowed bound states with definite total isospin and spin of
light degrees of freedom are given by \cite{GLM}
\begin{eqnarray}\nonumber
\left|Ra
s_lm\right>\left>_K\right.=\sqrt{\frac{\mbox{dim}R~\mbox{dim}s_l}{\mbox{dim}K}}\int_{SU(3)}
dg D_{ac}^{*(R)}(g)\\
\times\widehat{U}_I(g)\left|\Sigma_0\right)\widehat{U}_I(g)\left|Kk^\prime\right\}\left(\begin{array}{cc}I&s_l\\i_3
&m\end{array}\right|\left.\begin{array}{cc}K\\k^\prime\end{array}\right),
\end{eqnarray}
with the constraint
\begin{equation}
Y=Y_\Sigma+Y_k=1+Y_k,
\end{equation}
where the isospin projector $\widehat{U}_I$ is the exponential of
the isospin generators, $\left|\Sigma_0\right)$ are the states of
a soliton with $A=1$, $\left|Kk^\prime\right\}$ are the
representations of $K$, which describe the heavy (anti-)meson
states $\bar{q}$($q$) in this case, and $H_I^\prime$ and
$\widetilde{H}_I^\prime$ act on the products of
$\left|\Sigma_0\right)$ and $\left|Kk\right\}$ as
\begin{eqnarray}
H_I\left|\Sigma_0\right)\left|Kk\right\}=V_I(K^2-I_H^2-S^2_{lH})\left|\Sigma_0\right)\left|Kk\right\},\label{be}
\end{eqnarray}
$I$ and $Y$ are the isospin and hypercharge of the $SU(3)$ state
$c$ of the representation $R$, and $Y_K$ is the hypercharge of
$\bar{q}$($q$)-states with $K$. For the ground states of heavy
mesons and their antiparticles, $S_{lH}=1/2$ and $I_H=(0,1/2)$,
Eq.~(\ref{be}) gives
\begin{equation}
\left\{\begin{array}{ll}
        H_I=-3V_I/2,&K=0,\\
        H_I=0,&K=1/2,\\
        H_I=V_I/2,&K=1.
\end{array}\right.
\end{equation}

\par The $\bar{q}$-states and $q$-states transform as
the $\overline{3}$ and $3$ representations of the $SU(3)$ group
respectively. Under the $SU(2)\times SU(1)$ subgroup($I_{Y_K}$),
they are decomposed as $\bar{3}\rightarrow
\frac{1}{2}_{-\frac{1}{3}}\oplus 0_{\frac{2}{3}}$ and
$3\rightarrow \frac{1}{2}_{\frac{1}{3}}\oplus 0_{-\frac{2}{3}}$
and accordingly we have $\bar{q}$-states with
\{$K$=0,1;$Y_K=-\frac{1}{3}$\},
\{$K=\frac{1}{2}$;$Y_K=\frac{2}{3}$\} and $q$-states with
\{$K$=0,1;$Y_K=\frac{1}{3}$\},
\{$K=\frac{1}{2}$;$Y_K=-\frac{2}{3}$\}. Thus, the allowed $SU(3)$
representations for a baryon-heavy-meson bound states with spin
$s_l$, isospin $I$ and hypercharge $Y$ are those satisfying
\cite{GLM}
\begin{eqnarray}
\begin{array}{lll}
(a)&I=s_l,&Y=\frac{2}{3},\\
(b)&1\subset I\otimes s_l,&Y=\frac{2}{3},\\
(c)&\frac{1}{2}\subset I\otimes s_l,&Y=\frac{5}{3},\label{c1}
\end{array}
\end{eqnarray}
and baryon-anti-heavy-meson bound states should satisfy
\begin{eqnarray}
\begin{array}{lll}
(a)&I=s_l,&Y=\frac{4}{3},\\
(b)&1\subset I\otimes s_l,&Y=\frac{4}{3},\\
(c)&\frac{1}{2}\subset I\otimes s_l,&Y=\frac{1}{3}.\label{c2}
\end{array}
\end{eqnarray}

\section{The exotic states with $C=\pm1$}

The bound states of light solitons and the $\bar{q}$(q)-states can
give us the heavy baryon(bound states) states. We have mentioned
that $8=(1,1)$ and $10=(3,0)$ baryon multiplets are the lowest two
of the light solitons(baryons). And we have the following direct
products
\begin{eqnarray}
8\otimes\bar{3}&=&\bar{15}\oplus 6\oplus\bar{3},~s_l=(0,1),\label{8b3}\\
8\otimes 3&=&15\oplus\bar{6}\oplus 3,~s_l=(0,1),\\
10\otimes\bar{3}&=&24\oplus 6,~s_l=(1,2),\\
10\otimes 3&=&15\oplus 15^\prime,~s_l=(1,2).\label{103}
\end{eqnarray}
And we can calculate the binding energy($H_I$) for each $SU(3)$
heavy baryon multiplets given above. Among them, $\bar{3}$ with
$S=1/2$ and $6$ with $S=(1/2,3/2)$ give the $SU(3)$ baryon
multiplets in the lowest two $SU(4)$ multiplets of 20 dimensions
\cite{GLM}. In the follow, we will discuss the light flavor
$SU(3)$ $\bar{15}=(1,2)$ and 24=(3,1) multplets with $C=1$ and the
$SU(3)$ $\bar{6}=(0,2)$, 15=(2,1), 3=(1,0) and $15^\prime=(4,0)$
multiplets with $C=-1$. There exist exotic states that are
different from the predictions of the quark model and we will
investigate these exotic states. First of all, we use this
approach to give the average masses of $\Theta^0_c(uudd\bar{c})$
and $\Theta^*_c(uddd\bar{c},uudd\bar{c},uuud\bar{c})$ in two light
flavor case to show the equivalence of the two approaches of the
bound state method mentioned in the introduction.

\subsection{The Average Mass of $\Theta^0_c$ and $\Theta^*_c$}

In the $SU(2)$ case, from the product $N\otimes q$ states, we have
the follow states $\left|Is_l\right>$
\begin{equation}
\begin{array}{ll}
\left|00;\frac{1}{2}\right>,&H_I=3/2gF^\prime(0);\\
\left|01;\frac{1}{2}\right>,&H_I=-1/2gF^\prime(0);\\
\left|10;\frac{1}{2}\right>,&H_I=-1/2gF^\prime(0);\\
\left|11;\frac{1}{2}\right>,&H_I=3/2gF^\prime(0),-1/2gF^\prime(0).
\end{array}
\end{equation}
We first discuss one state, $\left|01\right>$, which describes the
baryons with the minimal quark content $\left|uudd\bar{c}\right>$
and charm number -1. However, it can give both exotic baryons with
$S=3/2$ and those with $S=1/2$ by the heavy quark spin symmetry,
thus, we use $\Theta^0_c$ and $\Theta^{0*}_c$ denoting the state
with $S=1/2$ and $S=3/2$ respectively and define
$\overline{m}_{\Theta^0_c}\equiv(4m_{\Theta^{0*}_c+}+2m_{\Theta^0_c})/6$.
There is only one parameter $gF^\prime(0)$ and by fitting the
$\Lambda_c$ mass, we have $gF^\prime(0)=419$ MeV \cite{GLM}. This
implies
\begin{equation}
\overline{m}_{\Theta_c^0}=m_N+m_H-gF^\prime(0)/2=2704~\mbox{MeV},
\end{equation}
where $m_H\equiv (3m_{D^*}+m_D)/4=1973$ MeV. To estimate the
average mass of $\Theta^*_c$ with $(I=1,s_l=1)$, we have to
include the state $\left|11;\frac{3}{2}\right>$ from direct
product $\Delta\otimes q$, and the matrix of interaction
hamiltonian between this two states is
\begin{eqnarray}
H_I=\frac{gF^\prime(0)}{6}\left(\begin{array}{cc}1&4\sqrt{2}\\4\sqrt{2}&5\end{array}\right)
+\left(\begin{array}{cc}0&0\\0&\Delta M\end{array}\right),
\end{eqnarray}
where $\Delta M$ is the mass difference between $N$ and $\Delta$
\cite{GLM}. Then, we get an exotic bound state with
$H_I=-\frac{gF^\prime(0)}{2}+\frac{1}{3}\Delta M$ and the average
mass is
\begin{equation}
\overline{m}_{\Theta_c^*}=m_N+m_H-gF^\prime(0)/2+\frac{1}{3}\Delta
M=2802~\mbox{MeV}.
\end{equation}
The results above are the same as those in Ref. \cite{OPM1}, which
indicates the equivalence of the two approaches.

\subsection{The Light Flavor $SU(3)$ Exotic Baryon Multiplets}
In (\ref{8b3})-(\ref{103}), exotic heavy baryon multiplets of the
$SU(3)$ representations are
\begin{equation}
\begin{array}{ll}
\bar{15}=(1,2),&24=(3,1),\\
15=(2,1),&\bar{6}=(0,2),\\
3=(1,0),&15^\prime=(4,0),
\end{array}
\end{equation}
which under the $SU(2)\times SU(1)$ subgroup decompose as
\begin{eqnarray}
\bar{15}&\rightarrow&\frac{1}{2}_\frac{5}{3}\oplus
0_\frac{2}{3}\oplus
1_\frac{2}{3}\oplus\frac{3}{2}_{-\frac{1}{3}}\oplus
\frac{1}{2}_{-\frac{1}{3}}\oplus 1_{-\frac{4}{3}},\\\nonumber
24&\rightarrow&\frac{3}{2}_\frac{5}{3}\oplus 1_\frac{2}{3}\oplus
2_\frac{2}{3}\oplus\frac{1}{2}_{-\frac{1}{3}}\oplus
\frac{3}{2}_{-\frac{1}{3}}\oplus 0_{-\frac{4}{3}}\\&&\oplus
1_{-\frac{4}{3}}\oplus \frac{1}{2}_{-\frac{7}{3}},\\
15&\rightarrow&\frac{1}{2}_{-\frac{5}{3}}\oplus
0_{-\frac{2}{3}}\oplus
1_{-\frac{2}{3}}\oplus\frac{3}{2}_{\frac{1}{3}}\oplus
\frac{1}{2}_{\frac{1}{3}}\oplus 1_{\frac{4}{3}},\\
\bar{6}&\rightarrow& 0_\frac{4}{3}\oplus
\frac{1}{2}_\frac{1}{3}\oplus 1_{-\frac{2}{3}},\\
3&\rightarrow&\frac{1}{2}_{\frac{1}{3}}\oplus 0_{-\frac{2}{3}},\\
15^\prime&\rightarrow&2_\frac{4}{3}\oplus\frac{3}{2}_\frac{1}{3}\oplus
1_{-\frac{2}{3}}\oplus\frac{1}{2}_{-\frac{5}{3}}\oplus
0_{-\frac{8}{3}}.
\end{eqnarray}

According to (\ref{c1}) and (\ref{c2}), there are the following
allowed exotic states denoted by $\{K,Y\}$
\begin{eqnarray}
\begin{array}{lll}
\bar{15}&&\\
s_l=0,&\{K=0,\frac{2}{3}\}, &H_I=-3/2gF^\prime(0);\\
&\{K=\frac{1}{2},\frac{5}{3}\},&H_I=0;\\
&\{K=1,\frac{2}{3}\}, &H_I=1/2gF^\prime(0);\\
s_l=1,&\{K=0,\frac{2}{3}\}, &H_I=-3/2gF^\prime(0);\\
&\{K=\frac{1}{2},\frac{5}{3}\},&H_I=0;\\
&\{K=1,\frac{2}{3}\}, &H_I=1/2gF^\prime(0);\\
&&  
\end{array}
\end{eqnarray}
\begin{eqnarray}
\begin{array}{lll}
24&&\\
s_l=1,&\{K=0,\frac{2}{3}\}, &H_I=-3/2gF^\prime(0);\\
&\{K=\frac{1}{2},\frac{5}{3}\},&H_I=0;\\
&\{K=1,\frac{2}{3}\}, &H_I=1/2gF^\prime(0);\\
s_l=2,&\{K=0,\frac{2}{3}\}, &H_I=-3/2gF^\prime(0);\\
&\{K=\frac{1}{2},\frac{5}{3}\},&H_I=0;\\
&\{K=1,\frac{2}{3}\}, &H_I=1/2gF^\prime(0);\\
{\bar{6}}&&\\
s_l=0,&\{K=0,\frac{4}{3}\}, &H_I=3/2gF^\prime(0);\\
&\{K=\frac{1}{2},\frac{1}{3}\}, &H_I=0;\\
s_l=1,&\{K=\frac{1}{2},\frac{1}{3}\}, &H_I=0;\\
&\{K=1,\frac{4}{3}\}, &H_I=-1/2gF^\prime(0);\\
&&\\
15&&\\
s_l=0,&\{K=\frac{1}{2},\frac{1}{3}\}, &H_I=0;\\
&\{K=1,\frac{4}{3}\}, &H_I=-1/2gF^\prime(0);\\\label{poss}
\end{array}
\end{eqnarray}
\begin{eqnarray}
\begin{array}{lll}
s_l=1,&\{K=0,\frac{4}{3}\}, &H_I=3/2gF^\prime(0);\\
&\{K=1,\frac{4}{3}\}, &H_I=-1/2gF^\prime(0);\\
&\{K=\frac{1}{2},\frac{1}{3}\}, &H_I=0;\\
s_l=2,&\{K=1,\frac{4}{3}\}, &H_I=-1/2gF^\prime(0);\\
&\{K=\frac{1}{2},\frac{1}{3}\}, &H_I=0;\\
&&\\
3&&\\
s_l=0,&\{K=\frac{1}{2},\frac{1}{3}\}, &H_I=0;\\
s_l=1,&\{K=\frac{1}{2},\frac{1}{3}\}, &H_I=0;\\
&&\\
15^\prime&&\\
s_l=1,&\{K=\frac{1}{2},\frac{1}{3}\},&H_I=0;\\
&\{K=1,\frac{4}{3}\}, &H_I=-1/2gF^\prime(0);\\
s_l=2,&\{K=0,\frac{4}{3}\}, &H_I=3/2gF^\prime(0);\\
&\{K=\frac{1}{2},\frac{1}{3}\},&H_I=0;\\
&\{K=1,\frac{4}{3}\}, &H_I=-1/2gF^\prime(0).\\
\end{array}
\end{eqnarray}

\section{Collective Coordination Quantization}

In Sec. IV, we see that $C=\pm1$ bound states are degenerated in
mass respectively. To obtain physical heavy baryons states with
correct spin and isospin, we have to calculate up to $1/N_c$ order
still in the heavy quark limits, \ie, $O(m_Q^0N_c^{-1})$.
Substituting the hedgehog Ansatz (\ref{heghog}) into (\ref{vmu})
and (\ref{amu}), we get
\begin{eqnarray}
\mathbf{V}&=&\left(\begin{array}{cc}i\frac{\sin^2(F(r)/2)}{r}(\mathbf{\tau}\times\widehat{\mathbf{r}})&0\\0&1\end{array}\right),
\nonumber\\
\mathbf{A}&=&\left(\begin{array}{cc}\frac{1}{2}\left[\left(\frac{\sin(F(r))}{r}-F^\prime(r)\right)(\mathbf{\tau}\cdot
\widehat{\mathbf{r}})\widehat{\mathbf{r}}-\frac{\sin(F(r))}{r}\mathbf{\tau}\right]&0\\0&1\end{array}\right),~~~~~
\nonumber
\end{eqnarray}
and from (\ref{L}) and (\ref{Ll}), we get
\begin{eqnarray}
V_{cl}&=&g\int
d^3x\mbox{Tr}(H\mathbf{\gamma}\gamma_5\cdot\mathbf{A}\bar{H}),\\
\widetilde{V}_{cl}&=&g\int
d^3x\mbox{Tr}(\bar{\widetilde{H}}\mathbf{\gamma}\gamma_5\cdot\mathbf{A}\widetilde{H}).
\end{eqnarray}

To evaluate $V_cl$ and $\widetilde{V}_{cl}$, we first find the
classical approximation to the wave functions of heavy mesons with
the general form
\begin{equation}
H=\widetilde{H}=\left(\begin{array}{c}\sqrt{\frac{1}{8\pi}}\widetilde{\phi}_{\pm}(\mathbf{\tau}\cdot
\widehat{\mathbf{r}})\chi h(r)\\0\end{array}\right),\label{hwave}
\end{equation}
where $\widetilde{\phi}_{\pm}$ are the two eigenstates of the
isospin of heavy mesons, the explicit formulae of $\chi$ are
defined in \cite{OPM1}, and the radio functions $h(r)$ of the
lowest energy eigenstate can be approximated by delta function,
which is normalized $\int r^2|h(r)|^2dr=1$. Accordingly. $H$ is
normalized as
\begin{equation}
-\int d^3x\mbox{Tr}(H\bar{H})=1,
\end{equation}
In (\ref{cllect}), we introduce the collective coordinate $A(t)$,
and the heavy meson fields transform simultaneously as
\begin{equation}
H=H_{bf}A(t)^+, \widetilde{H}=A(t)\widetilde{H}_{bf},\label{ch}
\end{equation}
where $H_{bf}$ is the heavy meson field in the isospin body-fixed
frame. Then, we obtain the Lagrangians as
\begin{eqnarray}\nonumber
L&=&-M_{cl}-V_{cl}-i\int
d^3x\mbox{Tr}(\partial_0H_{bf}\overline{H}_{bf})+\frac{1}{2}I_\pi
\sum\limits_{a=1}^3\omega^a\omega^a\\
&+&\frac{1}{2}I_K\sum\limits_{b=4}^7\omega^b\omega^b+\sum\limits_{c=1}^8I_{bf}^c\omega^c-\frac{N_cB}{2\sqrt{3}}\omega^8,\\\nonumber
\widetilde{L}&=&-M_{cl}-\widetilde{V}_{cl}-i\int
d^3x\mbox{Tr}(\overline{\widetilde{H}}_{bf}\partial_0\widetilde{H}_{bf})+\frac{1}{2}I_\pi
\sum\limits_{a=1}^3\omega^a\omega^a\\
&+&\frac{1}{2}I_K\sum\limits_{b=4}^7\omega^b\omega^b
+\sum\limits_{c=1}^8\widetilde{I}_{bf}^c\omega^c-\frac{N_cB}{2\sqrt{3}}\omega^8,
\end{eqnarray}
where $M_{cl}$ is the classical soliton mass, $I_\pi$ and $I_K$
are moments of inertia \cite{Guad}, $\omega^a$ are defined by
$A^\dagger\partial_0A=i\frac{1}{2}\omega^a\lambda^a$ and
$I_{bf}^a$ and $\widetilde{I}_{bf}^a$ are the rotation operator of
heavy mesons on the flavor $SU(3)$ group in the body fixed frame,
defined as
\begin{eqnarray}
I_{bf}^a=-\frac{1}{4}\int
d^3x[H_{bf}(\xi^\dagger_0\tau^a\xi_0+\xi_0\tau^a\xi^\dagger_0)\bar{H}_{bf}],\\
\widetilde{I}_{bf}^a=\frac{1}{4}\int
d^3x[\bar{\widetilde{H}}_{bf}(\xi^\dagger_0\tau^a\xi_0+\xi_0\tau^a\xi^\dagger_0)\widetilde{H}_{bf}].
\end{eqnarray}

By defining the canonical momenta conjugate to $\omega^a$ as
$R_a=-\frac{\delta L}{\delta\omega^a}$, we get the Hamiltonians
\begin{eqnarray}\nonumber
H&=&M_{cl}+V_{cl}+\frac{1}{2I_\pi}\sum\limits_{a=1}^3(R_a+I_{bf}^a)^2\\
&+&\frac{1}{2I_K}\sum\limits_{b=4}^7(R_b+I_{bf}^b)^2,\\\nonumber
\widetilde{H}&=&M_{cl}+\widetilde{V}_{cl}+\frac{1}{2I_\pi}\sum\limits_{a=1}^3(R_a+\widetilde{I}_{bf}^a)^2\\
&+&\frac{1}{2I_K}\sum\limits_{b=4}^7(R_b+\widetilde{I}_{bf}^b)^2,
\end{eqnarray}
and $\frac{2}{\sqrt{3}}R_8=\frac{2}{3}, \frac{4}{3}$ respectively.
Due to the embedding of $SU(2)$ heavy meson wave functions into
$SU(3)$ (\ref{hwave}), $I_{bf}^a$ and $\widetilde{I}_{bf}^a$ with
$a=4,5,6,7$ are zero and the mass formulae are as follow
\begin{eqnarray}\nonumber
M&=&M_{cl}+V_{cl}+\frac{1}{2I_K}C_2(SU(3))+\frac{1}{2}(\frac{1}{I_\pi}
-\frac{1}{I_K})j_{sol}(j_{sol}+1)\\
&-&\frac{1}{2I_K}R_8^2+\frac{1}{2I_\pi}(\mathbf{I}_{bf}^2+\mathbf{R}\cdot\mathbf{I}_{bf}),\label{Mc}
\end{eqnarray}
where $C_2(SU(3)=1/3(p^2+q^2+pq)+(p+q)$, $(p,q)$ denotes an
irreducible representation of the SU(3) group, $j_{sol}$ are the
angular momenta of soliton, and the last two terms reduce to
$SU(2)$ case in Ref.\cite{OPM}. And due to the the spin-grand-spin
transmutation \cite{OPM}, we have $s_l=j_{sol}+K$. Accordingly,
the wave functions are
\begin{equation}
    \Psi_{\nu\nu^\prime}^{(\mu)}(A)=\sqrt{\mbox{dim}(\mu)}D^{(\mu)}_{YII_3;Y_Rj_{sol}-j_{sol3}}(A)\label{psi2},
\end{equation}
with the constraints that $Y_R=2/3,4/3$ respectively for $C=\pm1$.
Then, these constraints mean $\bar{6}$-plet with $j_{sol}=0$,
$15$-plet with $j_{sol}=1$, $\bar{15}$-plet with $j_{sol}=(0,1)$,
and $24$-plet with $j_{sol}=(1,2)$. To get the mass splitting by
strange quark mass, we have to take into consideration the
following terms~\cite{wise}
\begin{eqnarray}\nonumber
&L_{SB}=\int d^3x\mbox{Tr}[\sigma_1H(\xi m^+_q\xi+\xi^\dagger
m_q\xi^\dagger)\bar{H}~~~~~~~\\
&~+\sigma_1^\prime\bar{H}_aH_a(\Sigma^\dagger
m^+_q+m_q\Sigma)-\Delta(\Sigma^\dagger
m^+_q+m_q\Sigma)],~~~~~\label{lsb}
\end{eqnarray}
with
\begin{eqnarray}
m_q=\left(\begin{array}{ccc}1&&\\&1&\\&&\widehat{m}\end{array}\right),
\end{eqnarray}
where $\widehat{m}=2m_s/(m_u+m_d)$ and the difference between
$m_u$ and $m_d$ is neglected. For both $C=\pm$, the symmetry
breaking hamiltonians are the same, and due to the orthogonality
of $H_{bf}$, the first two terms in (\ref{lsb}) give the same term
as the last after evaluating $H_{bf}$, and omitting the constant
terms, we have\cite{mss}
\begin{equation}
H_{SB}=\tau (1-D^{(8)}_{88}),
\end{equation}
where we use $\tau$ to denote all other parameters. Up to now, we
can use perturbation theory to calculate the mass splitting, the
results of the $\overline{6}$-plet and $15$-plet baryons are shown
in table 1, where the input masses of $\Theta^0_c$ and
$\Theta_c^*$ are the results in Sec. IV, and the Clebsch-Gordan
coefficients are listed in Appendix. To fix the parameter $\tau$,
we calculate the mass splitting between $\Xi_c$ and $\Sigma_c$,
whose experimental value is $\Delta
m=\bar{m}_{\Xi_c}-\bar{m}_{\Sigma_c}=89$ ~MeV on the average of
the degrees of freedom of spin. From the symmetry breaking
hamiltonian, we can get the splitting $\Delta m=3/20\tau$, and,
thus, we can estimate the masses of all the $\bar{6}$-plet and
15-plet baryons as shown in Table 1 and give the following mass
relations
\begin{eqnarray}
\bar{m}_{\Sigma_{\bar{c}}}-\bar{m}_{N_{\bar{c}}}
&=&\bar{m}_{N_{\bar{c}}}-\bar{m}_{\Theta^0_c}=2(\bar{m}_{\Xi_c}-\bar{m}_{\Sigma_c});\\
\bar{m}_{\Sigma^*_{\bar{c}}}+\bar{m}_{\Theta^*_c}&=&\bar{m}_{\Lambda_{\bar{c}}}+\bar{m}_{N^*_{\bar{c}}};\\
\bar{m}_{\Xi_{\bar{c}}}+\bar{m}_{\Theta^*_c}&=&\bar{m}_{\Delta_{\bar{c}}}+\bar{m}_{N^*_{\bar{c}}};\\
\bar{m}_{\Sigma^*_{\bar{c}}}-\bar{m}_{\Theta^*_c}&=&2(\bar{m}_{\Delta_{\bar{c}}}-\bar{m}_{N^*_{\bar{c}}});\\
\bar{m}_{\Sigma^*_{\bar{c}}}-\bar{m}_{\Theta^*_c}&=&2(\bar{m}_{\Xi_{\bar{c}}}-\bar{m}_{\Lambda_{\bar{c}}});\\
\bar{m}_{\Delta_{\bar{c}}}+\bar{m}_{\Xi_{\bar{c}}}&=&2\bar{m}_{\Sigma^*_{\bar{c}}}.
\end{eqnarray}
\begin{center}
\begin{tabular}{c|cccc}
\multicolumn{4}{c}{Table 1. Masses of the $\overline{6}$-plet
 and 15-plet baryons}\\\hline\\
 Baryon States& $I(S_l^P)$ &$\left<H_{SB}\right>$
& M~(Mev)\\
\hline
$\Theta_c^0$&0$(1^+)$&$\frac{3}{5}\tau$&2704\\
$N_{\bar{c}}$&$\frac{1}{2}(1^+)$&$\frac{9}{10}\tau$&2882\\
$\Sigma_{\bar{c}}$&$1(1^+)$&$\frac{6}{5}\tau$&3060
\\\hline
$\Theta_c^{*}$&1$((1^+)$&$\frac{2}{3}\tau$&2802\\
$\Delta_{\bar{c}}$&$\frac{3}{2}(1^+)$&$\frac{13}{12}\tau$&3049\\
$N_{\bar{c}}^{*}$&$\frac{1}{2}(1^+)$&$\frac{5}{6}\tau$&2901\\
$\Sigma_{\bar{c}}^*$&1$(1^+)$&$\frac{7}{6}\tau$&3099\\
$\Lambda_{\bar{c}}$&0$(1^+)$&$\tau$&3000\\
$\Xi_{\bar{c}}$&$\frac{1}{2}(1^+)$&$\frac{5}{4}\tau$&3148\\\hline
\end{tabular}
\end{center}

\section{Summary and Discussion}

In this paper, we first illustrate the $SU(3)$ exotic baryon
multiplets as the sub-multiplets of $SU(4)$ baryon multiplets
predicted from the generalized flavor $SU(4)$
collective-coordinate quantization in Figs.\ref{fig1} and
\ref{fig2}. Then, we generalize the approach in \cite{GLM} to
heavy baryons with anti-charm quark, and calculate the average
masses of the states $\Theta^0_c$ and $\Theta^{*}_c$ in the light
flavor $SU(2)$ case and investigate the binding energy of the
heavy baryon multiplets with exotic states up to leading order of
$1/N_c$ under the light flavor $SU(3)$ symmetry and the heavy
quark limit. Then, by collective coordinate quantization, we give
the mass spectra of the $\bar{6}$-plet and 15-plet baryons. In the
light flavor $SU(2)$ case, we calculate the average masses of
$\Theta^0_c$ and $\Theta^*_c$, which are estimated to be 2704~MeV
and 2802~MeV respectively, and in the picture of the light flavor
$SU(3)$ symmetry, the $\Theta^0_c$ and $\Theta^*_c$ states belong
to the $\bar{6}$-plets and 15-plet with $S=(1/2,3/2)$. These
masses are the same as that in \cite{OPM1}, but in that approach
they also predicted the existence of $\Theta^0_c$ and $\Theta^*_c$
with negative parity. Calculations in the light flavor $SU(3)$
case show that there exist $\bar{15}$-plets and 24-plets with
$C=1$ and $\bar{6}$-plet and $15$-plets with $C=-1$ bounded in the
bound state approach, qualitatively consistent with the results in
Figs.~\ref{fig1} and \ref{fig2}. However, in this approach $SU(3)$
baryon multiplets $\bar{6}$-plet with $S=3/2$ and
$15^\prime$-plets with $S=(1/2,3/2),(3/2,5/2)$ are also bounded.
We find that these bounded states with $C=1$ and those with $C=-1$
have equal binding energy respectively, this is because
(\ref{hip}) and (\ref{whip}) are only up to the leading order of
$1/N_c$. Then, we quantize the chiral $SU(3)$ Lagrangians and
calculate the splitting masses of the $\bar{6}$-plet  and 15-plet
baryons and give some relations between them, which may be of
greater significance qualitatively than quantitatively because in
the case of $SU(3)$ chiral symmetry with only light baryons, such
a way to deal with symmetry breaking term does not agree with
experiments very well \cite{Chem}.
\par
In summary, both bound state approaches and the generalized
$SU(N_f)$ collective-coordinate quantization approach give us
exotic baryon states with $C=\pm1$, and the search of $\Theta^0_c$
are expected further to verify the validation of the
generalization of soliton picture to study exotic baryon states.
\par

\section*{ACKNOWLEDGMENT}

We are grateful for discussions with Han-Qing Zheng. This work is
partially supported by National Natural Science Foundation of
China under Grant Numbers 10025523 and 90103007.

\section*{APPENDIX: The Clebsch-Gordan Coefficients used in the paper}
Because the Clebsch-Gordan coefficients used in this paper are
seldom used in physics before, we list the Clebsch-Gordan
Coefficients in this appendix. To be convenient, we define
\begin{eqnarray}
\mu^{(\lambda)}_{(\nu)}=\sum\limits_{\nu_1,\nu_2}
\left(\begin{array}{ccc}\mu_1&\mu_2&\mu\\\nu_1&\nu_2&\nu\end{array}\right)\mu_{1(\nu_1)}\mu_{2(\nu_2)},
\end{eqnarray}
where $\mu_{(\nu)}^{\lambda}$ denote the eigenstates of the
representation $\mu$ contained in the direct sum of $\mu_1$ and
$\mu_2$, whose enginstates are $\mu_{1(\nu_1)}$ and
$\mu_{2(\nu_2)}$ respectively, $\lambda$ is used to distinguish
identical but independent representations which are all contained
in $\mu_1\otimes\mu_2$, $\nu$,$\nu_1$ and $\nu_2$ denote quantum
number $(YII_3)$ collectively, and
$\left(\begin{array}{ccc}\mu_1&\mu_2&\mu\\\nu_1&\nu_2&\nu\end{array}\right)$
are the $SU(3)$ Clebsch-Gordan coefficients. For the product
$\bar{6}\otimes8=3\oplus\bar{6}\oplus15\oplus\bar{24}$, we give
the decomposition of $\bar{6}$ as follows
\begin{widetext}
\begin{eqnarray}\nonumber
\bar{6}_{(\frac{4}{3},0,0)}&=&\sqrt{\frac{2}{5}}\bar{6}_{(\frac{4}{3},0,0)}8_{(0,0,0)}
-\sqrt{\frac{3}{10}}\bar{6}_{(\frac{1}{3},\frac{1}{2},\frac{1}{2})}8_{(1,\frac{1}{2},-\frac{1}{2})}
+\sqrt{\frac{3}{10}}\bar{6}_{(\frac{1}{3},\frac{1}{2},-\frac{1}{2})}8_{(1,\frac{1}{2},\frac{1}{2})},\\\nonumber
\bar{6}_{(\frac{1}{3},\frac{1}{2},\frac{1}{2})}&=&\sqrt{\frac{3}{10}}\bar{6}_{(\frac{4}{3},0,0)}8_{(-1,\frac{1}{2},\frac{1}{2})}
-\sqrt{\frac{3}{40}}\bar{6}_{(\frac{1}{3},\frac{1}{2},\frac{1}{2})}8_{(0,1,0)}
+\sqrt{\frac{1}{40}}\bar{6}_{(\frac{1}{3},\frac{1}{2},\frac{1}{2})}8_{(0,0,0)}
+\sqrt{\frac{3}{20}}\bar{6}_{(\frac{1}{3},\frac{1}{2},-\frac{1}{2})}8_{(0,1,1)}\\\nonumber
&-&\sqrt{\frac{3}{10}}\bar{6}_{(-\frac{2}{3},1,1)}8_{(1,\frac{1}{2},-\frac{1}{2})}
+\sqrt{\frac{3}{20}}\bar{6}_{(-\frac{2}{3},1,0)}8_{(1,\frac{1}{2},\frac{1}{2})},\\\nonumber
\bar{6}_{(\frac{1}{3},\frac{1}{2},-\frac{1}{2})}&=&\sqrt{\frac{3}{10}}\bar{6}_{(\frac{4}{3},0,0)}8_{(-1,\frac{1}{2},-\frac{1}{2})}
-\sqrt{\frac{3}{20}}\bar{6}_{(\frac{1}{3},\frac{1}{2},\frac{1}{2})}8_{(0,1,-1)}
+\sqrt{\frac{1}{40}}\bar{6}_{(\frac{1}{3},\frac{1}{2},-\frac{1}{2})}8_{(0,0,0)}
+\sqrt{\frac{3}{40}}\bar{6}_{(\frac{1}{3},\frac{1}{2},-\frac{1}{2})}8_{(0,1,0)}\\\nonumber
&-&\sqrt{\frac{3}{20}}\bar{6}_{(-\frac{2}{3},1,0)}8_{(1,\frac{1}{2},-\frac{1}{2})}
+\sqrt{\frac{3}{10}}\bar{6}_{(-\frac{2}{3},1,-1)}8_{(1,\frac{1}{2},\frac{1}{2})},\\\nonumber
\bar{6}_{(-\frac{2}{3},1,1)}&=&\sqrt{\frac{3}{10}}\bar{6}_{(\frac{1}{3},\frac{1}{2},\frac{1}{2})}8_{(-1,\frac{1}{2},\frac{1}{2})}
-\sqrt{\frac{3}{10}}\bar{6}_{(-\frac{2}{3},1,1)}8_{(0,1,0)}
-\sqrt{\frac{1}{10}}\bar{6}_{(-\frac{2}{3},1,1)}8_{(0,0,0)}
+\sqrt{\frac{3}{10}}\bar{6}_{(-\frac{2}{3},1,0)}8_{(0,1,1)},\\\nonumber
\bar{6}_{(-\frac{2}{3},1,0)}&=&\sqrt{\frac{3}{20}}\bar{6}_{(\frac{1}{3},\frac{1}{2},\frac{1}{2})}8_{(-1,\frac{1}{2},-\frac{1}{2})}
+\sqrt{\frac{3}{20}}\bar{6}_{(\frac{1}{3},\frac{1}{2},-\frac{1}{2})}8_{(-1,\frac{1}{2},\frac{1}{2})}
-\sqrt{\frac{3}{10}}\bar{6}_{(-\frac{2}{3},1,1)}8_{(0,1,-1)}
-\sqrt{\frac{1}{10}}\bar{6}_{(-\frac{2}{3},1,0)}8_{(0,0,0)}\\\nonumber
&+&\sqrt{\frac{3}{10}}\bar{6}_{(-\frac{2}{3},1,-1)}8_{(0,1,1)},\\\nonumber
\bar{6}_{(-\frac{2}{3},1,-1)}&=&\sqrt{\frac{3}{10}}\bar{6}_{(\frac{1}{3},\frac{1}{2},-\frac{1}{2})}8_{(-1,\frac{1}{2},-\frac{1}{2})}
-\sqrt{\frac{3}{10}}\bar{6}_{(-\frac{2}{3},1,0)}8_{(0,1,-1)}
+\sqrt{\frac{3}{10}}\bar{6}_{(-\frac{2}{3},1,-1)}8_{(0,1,0)}
-\sqrt{\frac{1}{10}}\bar{6}_{(-\frac{2}{3},1,-1)}8_{(0,0,0)}.
\end{eqnarray}

Similarly, in the product
$15\otimes8=42\oplus\bar{24}\oplus15^\prime\oplus15^{(1)}\oplus15^{(2)}\oplus\bar{6}\oplus3$,
the representations $15^{(1)}$ and $15^{(2)}$ are given as

\begin{eqnarray}\nonumber
15^{(1)}_{(\frac{4}{3},1,1)}&=&
\frac{\sqrt{21}}{14}15_{(\frac{4}{3},1,1)}8_{(0,1,0)}
+\frac{\sqrt{7}}{7}15_{(\frac{4}{3},1,1)}8_{(0,0,0)}
-\frac{\sqrt{21}}{14}15_{(\frac{4}{3},1,0)}8_{(0,1,1)}
-\frac{3\sqrt{42}}{28}15_{(\frac{1}{3},\frac{3}{2},\frac{3}{2})}8_{(1,\frac{1}{2},-\frac{1}{2})}\\\nonumber
&+&\frac{3\sqrt{14}}{28}15_{(\frac{1}{3},\frac{3}{2},\frac{1}{2})}8_{(1,\frac{1}{2},\frac{1}{2})},\\\nonumber
15^{(1)}_{(\frac{4}{3},1,0)}&=&
\frac{\sqrt{21}}{14}15_{(\frac{4}{3},1,1)}8_{(0,1,-1)}
+\frac{\sqrt{7}}{7}15_{(\frac{4}{3},1,0)}8_{(0,0,0)}
-\frac{\sqrt{21}}{14}15_{(\frac{4}{3},1,-1)}8_{(0,1,1)}
-\frac{3\sqrt{7}}{14}15_{(\frac{1}{3},\frac{3}{2},\frac{1}{2})}8_{(1,\frac{1}{2},-\frac{1}{2})}\\\nonumber
&+&\frac{3\sqrt{7}}{14}15_{(\frac{1}{3},\frac{3}{2},-\frac{1}{2})}8_{(1,\frac{1}{2},\frac{1}{2})},\\\nonumber
15^{(1)}_{(\frac{4}{3},1,-1)}&=&
\frac{\sqrt{21}}{14}15_{(\frac{4}{3},1,0)}8_{(0,1,-1)}
-\frac{\sqrt{21}}{14}15_{(\frac{4}{3},1,-1)}8_{(0,1,0)}
+\frac{\sqrt{7}}{7}15_{(\frac{4}{3},1,-1)}8_{(0,0,0)}
-\frac{3\sqrt{14}}{28}15_{(\frac{1}{3},\frac{3}{2},-\frac{1}{2})}8_{(1,\frac{1}{2},-\frac{1}{2})}\\\nonumber
&+&\frac{3\sqrt{42}}{28}15_{(\frac{1}{3},\frac{3}{2},-\frac{3}{2})}8_{(1,\frac{1}{2},\frac{1}{2})},\\\nonumber
15^{(1)}_{(\frac{1}{3},\frac{3}{2},\frac{3}{2})}&=&
\frac{3\sqrt{42}}{28}15_{(\frac{4}{3},1,1)}8_{(-1,\frac{1}{2},\frac{1}{2})}
-\frac{\sqrt{21}}{28}15_{(\frac{1}{3},\frac{3}{2},\frac{3}{2})}8_{(0,1,0)}
-\frac{5\sqrt{7}}{28}15_{(\frac{1}{3},\frac{3}{2},\frac{3}{2})}8_{(0,0,0)}
+\frac{\sqrt{14}}{28}15_{(\frac{1}{3},\frac{3}{2},\frac{1}{2})}8_{(0,1,1)}\\\nonumber
&-&\frac{\sqrt{7}}{7}15_{(\frac{1}{3},\frac{1}{2},\frac{1}{2})}8_{(0,1,1)}
+\frac{\sqrt{21}}{14}15_{(-\frac{2}{3},1,1)}8_{(1,\frac{1}{2},\frac{1}{2})},\\\nonumber
15^{(1)}_{(\frac{1}{3},\frac{3}{2},\frac{1}{2})}&=&
\frac{3\sqrt{14}}{28}15_{(\frac{4}{3},1,1)}8_{(-1,\frac{1}{2},-\frac{1}{2})}
+\frac{3\sqrt{7}}{14}15_{(\frac{4}{3},1,0)}8_{(-1,\frac{1}{2},\frac{1}{2})}
-\frac{\sqrt{14}}{28}15_{(\frac{1}{3},\frac{3}{2},\frac{3}{2})}8_{(0,1,-1)}
-\frac{\sqrt{21}}{84}15_{(\frac{1}{3},\frac{3}{2},\frac{1}{2})}8_{(0,1,0)}\\\nonumber
&-&\frac{5\sqrt{7}}{28}15_{(\frac{1}{3},\frac{3}{2},\frac{1}{2})}8_{(0,0,0)}
-\frac{\sqrt{42}}{21}15_{(\frac{1}{3},\frac{1}{2},\frac{1}{2})}8_{(0,1,0)}
+\frac{\sqrt{42}}{42}15_{(\frac{1}{3},\frac{3}{2},-\frac{1}{2})}8_{(0,1,1)}
-\frac{\sqrt{21}}{21}15_{(\frac{1}{3},\frac{1}{2},-\frac{1}{2})}8_{(0,1,1)}\\\nonumber
&+&\frac{\sqrt{7}}{14}15_{(-\frac{2}{3},1,1)}8_{(1,\frac{1}{2},-\frac{1}{2})}
+\frac{\sqrt{14}}{14}15_{(-\frac{2}{3},1,0)}8_{(1,\frac{1}{2},\frac{1}{2})},\\\nonumber
15^{(1)}_{(\frac{1}{3},\frac{3}{2},-\frac{1}{2})}&=&
\frac{3\sqrt{7}}{14}15_{(\frac{4}{3},1,0)}8_{(-1,\frac{1}{2},-\frac{1}{2})}
-\frac{\sqrt{42}}{42}15_{(\frac{1}{3},\frac{3}{2},\frac{1}{2})}8_{(0,1,-1)}
-\frac{\sqrt{21}}{21}15_{(\frac{1}{3},\frac{1}{2},\frac{1}{2})}8_{(0,1,-1)}
+\frac{3\sqrt{14}}{28}15_{(\frac{4}{3},1,-1)}8_{(-1,\frac{1}{2},\frac{1}{2})}\\\nonumber
&+&\frac{\sqrt{21}}{84}15_{(\frac{1}{3},\frac{3}{2},-\frac{1}{2})}8_{(0,1,0)}
-\frac{5\sqrt{7}}{28}15_{(\frac{1}{3},\frac{3}{2},-\frac{1}{2})}8_{(0,0,0)}
-\frac{\sqrt{42}}{21}15_{(\frac{1}{3},\frac{1}{2},-\frac{1}{2})}8_{(0,1,0)}
+\frac{\sqrt{14}}{28}15_{(\frac{1}{3},\frac{3}{2},-\frac{3}{2})}8_{(0,1,1)}\\\nonumber
&+&\frac{\sqrt{14}}{14}15_{(-\frac{2}{3},1,0)}8_{(1,\frac{1}{2},-\frac{1}{2})}
+\frac{\sqrt{7}}{14}15_{(-\frac{2}{3},1,-1)}8_{(1,\frac{1}{2},\frac{1}{2})},\\\nonumber
15^{(1)}_{(\frac{1}{3},\frac{3}{2},-\frac{3}{2})}&=&
\frac{3\sqrt{42}}{28}15_{(\frac{4}{3},1,-1)}8_{(-1,\frac{1}{2},-\frac{1}{2})}
-\frac{\sqrt{14}}{28}15_{(\frac{1}{3},\frac{3}{2},-\frac{1}{2})}8_{(0,1,-1)}
-\frac{\sqrt{7}}{7}15_{(\frac{1}{3},\frac{1}{2},-\frac{1}{2})}8_{(0,1,-1)}
+\frac{\sqrt{21}}{28}15_{(\frac{1}{3},\frac{3}{2},-\frac{3}{2})}8_{(0,1,0)}\\\nonumber
&-&\frac{5\sqrt{7}}{28}15_{(\frac{1}{3},\frac{3}{2},-\frac{3}{2})}8_{(0,0,0)}
+\frac{\sqrt{21}}{14}15_{(-\frac{2}{3},1,-1)}8_{(1,\frac{1}{2},-\frac{1}{2})},\\\nonumber
15^{(1)}_{(\frac{1}{3},\frac{1}{2},\frac{1}{2})}&=&
\frac{\sqrt{7}}{7}15_{(\frac{1}{3},\frac{3}{2},\frac{3}{2})}8_{(0,1,-1)}
-\frac{\sqrt{42}}{21}15_{(\frac{1}{3},\frac{3}{2},\frac{1}{2})}8_{(0,1,0)}
+\frac{\sqrt{21}}{21}15_{(\frac{1}{3},\frac{1}{2},\frac{1}{2})}8_{(0,1,0)}
+\frac{\sqrt{7}}{7}15_{(\frac{1}{3},\frac{1}{2},\frac{1}{2})}8_{(0,0,0)}\\\nonumber
&+&\frac{\sqrt{21}}{21}15_{(\frac{1}{3},\frac{3}{2},-\frac{1}{2})}8_{(0,1,1)}
-\frac{\sqrt{42}}{21}15_{(\frac{1}{3},\frac{1}{2},-\frac{1}{2})}8_{(0,1,1)}
-\frac{\sqrt{14}}{7}15_{(-\frac{2}{3},1,1)}8_{(1,\frac{1}{2},-\frac{1}{2})}
+\frac{\sqrt{7}}{7}15_{(-\frac{2}{3},1,0)}8_{(1,\frac{1}{2},\frac{1}{2})},\\\nonumber
15^{(1)}_{(\frac{1}{3},\frac{1}{2},-\frac{1}{2})}&=&
\frac{\sqrt{21}}{21}15_{(\frac{1}{3},\frac{3}{2},\frac{1}{2})}8_{(0,1,-1)}
+\frac{\sqrt{42}}{21}15_{(\frac{2}{3},\frac{1}{2},\frac{1}{2})}8_{(0,1,-1)}
-\frac{\sqrt{42}}{21}15_{(\frac{1}{3},\frac{3}{2},-\frac{1}{2})}8_{(0,1,0)}
-\frac{\sqrt{21}}{21}15_{(\frac{1}{3},\frac{1}{2},-\frac{1}{2})}8_{(0,1,0)}\\\nonumber
&+&\frac{\sqrt{7}}{7}15_{(\frac{1}{3},\frac{1}{2},-\frac{1}{2})}8_{(0,0,0)}
+\frac{\sqrt{7}}{7}15_{(\frac{1}{3},\frac{3}{2},-\frac{3}{2})}8_{(0,1,1)}
-\frac{\sqrt{7}}{7}15_{(-\frac{2}{3},1,0)}8_{(1,\frac{1}{2},-\frac{1}{2})}
+\frac{\sqrt{14}}{7}15_{(-\frac{2}{3},1,-1)}8_{(1,\frac{1}{2},\frac{1}{2})},\\\nonumber
15^{(1)}_{(-\frac{2}{3},1,1)}&=&
\frac{\sqrt{21}}{14}15_{(\frac{1}{3},\frac{3}{2},\frac{3}{2})}8_{(-1,\frac{1}{2},-\frac{1}{2})}
-\frac{\sqrt{7}}{14}15_{(\frac{1}{3},\frac{3}{2},\frac{1}{2})}8_{(-1,\frac{1}{2},\frac{1}{2})}
+\frac{\sqrt{14}}{7}15_{(\frac{1}{3},\frac{1}{2},\frac{1}{2})}8_{(-1,\frac{1}{2},\frac{1}{2})}
-\frac{\sqrt{21}}{14}15_{(-\frac{2}{3},1,1)}8_{(0,1,0)}\\\nonumber
&-&\frac{\sqrt{7}}{14}15_{(-\frac{2}{3},1,1)}8_{(0,0,0)}
+\frac{\sqrt{21}}{14}15_{(-\frac{2}{3},1,0)}8_{(0,1,1)}
-\frac{\sqrt{42}}{14}15_{(-\frac{2}{3},0,0)}8_{(0,1,1)}
+\frac{\sqrt{21}}{14}15_{(-\frac{5}{3},\frac{1}{2},\frac{1}{2})}8_{(1,\frac{1}{2},\frac{1}{2})},\\\nonumber
\end{eqnarray}
\begin{eqnarray}\nonumber
15^{(1)}_{(-\frac{2}{3},1,0)}&=&
\frac{\sqrt{14}}{14}15_{(\frac{1}{3},\frac{3}{2},\frac{1}{2})}8_{(-1,\frac{1}{2},-\frac{1}{2})}
-\frac{\sqrt{14}}{14}15_{(\frac{1}{3},\frac{3}{2},-\frac{1}{2})}8_{(-1,\frac{1}{2},\frac{1}{2})}
+\frac{\sqrt{7}}{7}15_{(\frac{1}{3},\frac{1}{2},-\frac{1}{2})}8_{(-1,\frac{1}{2},\frac{1}{2})}
+\frac{\sqrt{7}}{7}15_{(\frac{1}{3},\frac{1}{2},\frac{1}{2})}8_{(-1,\frac{1}{2},-\frac{1}{2})}\\\nonumber
&-&\frac{\sqrt{28}}{28}15_{(-\frac{2}{3},1,0}8_{(0,0,0)}
-\frac{\sqrt{42}}{14}15_{(-\frac{2}{3},0,0)}8_{(0,1,0)}
+\frac{\sqrt{42}}{28}15_{(-\frac{5}{3},\frac{1}{2},\frac{1}{2})}8_{(1,\frac{1}{2},-\frac{1}{2})}
+\frac{\sqrt{21}}{14}15_{(-\frac{2}{3},1,-1)}8_{(0,1,1)}\\\nonumber
&+&\frac{\sqrt{42}}{28}15_{(-\frac{5}{3},\frac{1}{2},-\frac{1}{2})}8_{(1,\frac{1}{2},\frac{1}{2})}
-\frac{\sqrt{21}}{14}15_{(-\frac{2}{3},1,1)}8_{(0,1,-1)},\\\nonumber
15^{(1)}_{(-\frac{2}{3},1,-1)}&=&
\frac{\sqrt{28}}{28}15_{(\frac{1}{3},\frac{3}{2},-\frac{1}{2})}8_{(-1,\frac{1}{2},-\frac{1}{2})}
+\frac{\sqrt{14}}{7}15_{(\frac{1}{3},\frac{1}{2},-\frac{1}{2})}8_{(-1,\frac{1}{2},-\frac{1}{2})}
-\frac{\sqrt{21}}{14}15_{(-\frac{2}{3},1,0)}8_{(0,1,-1)}
-\frac{\sqrt{42}}{14}15_{(-\frac{2}{3},0,0)}8_{(0,1,-1)}\\\nonumber
&-&\frac{\sqrt{21}}{14}15_{(\frac{1}{3},\frac{3}{2},-\frac{3}{2})}8_{(-1,\frac{1}{2},\frac{1}{2})}
+\frac{\sqrt{21}}{14}15_{(-\frac{2}{3},1,-1)}8_{(0,1,0)}
-\frac{\sqrt{28}}{28}15_{(-\frac{2}{3},1,-1)}8_{(0,0,0)}
+\frac{\sqrt{21}}{14}15_{(-\frac{5}{3},\frac{1}{2},-\frac{1}{2})}8_{(1,\frac{1}{2},-\frac{1}{2})},\\\nonumber
15^{(1)}_{(-\frac{2}{3},0,0)}&=&
\frac{\sqrt{42}}{14}15_{(-\frac{2}{3},1,1)}8_{(0,1,-1)}
-\frac{\sqrt{42}}{14}15_{(-\frac{2}{3},1,0)}8_{(0,1,0)}
+\frac{\sqrt{7}}{7}15_{(-\frac{2}{3},0,0))}8_{(0,0,0)}
-\frac{\sqrt{21}}{14}15_{(-\frac{5}{3},\frac{1}{2},\frac{1}{2})}8_{(1,\frac{1}{2},-\frac{1}{2})}\\\nonumber
&+&\frac{\sqrt{42}}{14}15_{(-\frac{2}{3},1,-1)}8_{(0,1,1)}
+\frac{\sqrt{21}}{14}15_{(-\frac{5}{3},\frac{1}{2},-\frac{1}{2})}8_{(1,\frac{1}{2},\frac{1}{2})},\\\nonumber
15^{(1)}_{(-\frac{5}{3},\frac{1}{2},\frac{1}{2})}&=&
\frac{\sqrt{21}}{14}15_{(-\frac{2}{3},1,1)}8_{(-1,\frac{1}{2},-\frac{1}{2})}
-\frac{\sqrt{42}}{28}15_{(-\frac{2}{3},1,0)}8_{(-1,\frac{1}{2},\frac{1}{2})}
+\frac{\sqrt{21}}{14}15_{(-\frac{2}{3},0,0)}8_{(-1,\frac{1}{2},\frac{1}{2})}
-\frac{3\sqrt{21}}{28}15_{(-\frac{5}{3},\frac{1}{2},\frac{1}{2})}8_{(0,1,0)}\\\nonumber
&+&\frac{\sqrt{7}}{28}15_{(-\frac{5}{3},\frac{1}{2},\frac{1}{2})}8_{(0,0,0)}
+\frac{3\sqrt{42}}{28}15_{(-\frac{5}{3},\frac{1}{2},-\frac{1}{2})}8_{(0,1,1)},\\\nonumber
15^{(1)}_{(-\frac{5}{3},\frac{1}{2},-\frac{1}{2})}&=&
\frac{\sqrt{42}}{28}15_{(-\frac{2}{3},1,0)}8_{(-1,\frac{1}{2},-\frac{1}{2})}
+\frac{\sqrt{21}}{14}15_{(-\frac{2}{3},0,0)}8_{(-1,\frac{1}{2},-\frac{1}{2})}
-\frac{\sqrt{21}}{14}15_{(-\frac{2}{3},1,-1)}8_{(-1,\frac{1}{2},\frac{1}{2})}
-\frac{3\sqrt{42}}{28}15_{(-\frac{5}{3},\frac{1}{2},\frac{1}{2})}8_{(0,1,-1)}\\\nonumber
&+&\frac{3\sqrt{21}}{28}15_{(-\frac{5}{3},\frac{1}{2},-\frac{1}{2})}8_{(0,1,0)}
+\frac{\sqrt{7}}{28}15_{(-\frac{5}{3},\frac{1}{2},-\frac{1}{2})}8_{(0,0,0)}.\\\nonumber
\end{eqnarray}


\begin{eqnarray}\nonumber
15^{(2)}_{(\frac{4}{3},1,1)}&=&
\frac{4\sqrt{7}}{21}15_{(\frac{4}{3},1,1)}8_{(0,1,0)}
-\frac{2\sqrt{21}}{21}15_{(\frac{4}{3},1,1)}8_{(0,0,0)}
-\frac{4\sqrt{7}}{21}15_{(\frac{4}{3},1,0)}8_{(0,1,1)}
+\frac{\sqrt{14}}{21}15_{(\frac{1}{3},\frac{3}{2},\frac{3}{2})}8_{(1,\frac{1}{2},-\frac{1}{2})}\\\nonumber
&-&\frac{\sqrt{42}}{63}15_{(\frac{1}{3},\frac{3}{2},\frac{1}{2})}8_{(1,\frac{1}{2},\frac{1}{2})}
+\frac{\sqrt{21}}{9}15_{(\frac{1}{3},\frac{1}{2},\frac{1}{2})}8_{(1,\frac{1}{2},\frac{1}{2})},\\\nonumber
15^{(2)}_{(\frac{4}{3},1,0)}&=&
\frac{4\sqrt{7}}{21}15_{(\frac{4}{3},1,1)}8_{(0,1,-1)}
-\frac{2\sqrt{21}}{21}15_{(\frac{4}{3},1,0)}8_{(0,0,0)}
-\frac{4\sqrt{7}}{21}15_{(\frac{4}{3},1,-1)}8_{(0,1,1)}
+\frac{2\sqrt{21}}{63}15_{(\frac{1}{3},\frac{3}{2},\frac{1}{2})}8_{(1,\frac{1}{2},-\frac{1}{2})}\\\nonumber
&+&\frac{\sqrt{42}}{18}15_{(\frac{1}{3},\frac{1}{2},\frac{1}{2})}8_{(1,\frac{1}{2},-\frac{1}{2})}
-\frac{2\sqrt{21}}{63}15_{(\frac{1}{3},\frac{3}{2},-\frac{1}{2})}8_{(1,\frac{1}{2},\frac{1}{2})}
+\frac{\sqrt{42}}{18}15_{(\frac{1}{3},\frac{1}{2},-\frac{1}{2})}8_{(1,\frac{1}{2},\frac{1}{2})},\\\nonumber
15^{(2)}_{(\frac{4}{3},1,-1)}&=&
\frac{4\sqrt{7}}{21}15_{(\frac{4}{3},1,0)}8_{(0,1,-1)}
-\frac{4\sqrt{7}}{21}15_{(\frac{4}{3},1,-1)}8_{(0,1,0)}
-\frac{2\sqrt{21}}{21}15_{(\frac{4}{3},1,-1)}8_{(0,0,0)}
+\frac{\sqrt{42}}{63}15_{(\frac{1}{3},\frac{3}{2},-\frac{1}{2})}8_{(1,\frac{1}{2},-\frac{1}{2})}\\\nonumber
&+&\frac{\sqrt{21}}{9}15_{(\frac{1}{3},\frac{1}{2},-\frac{1}{2})}8_{(1,\frac{1}{2},-\frac{1}{2})}
-\frac{\sqrt{14}}{21}15_{(\frac{1}{3},\frac{3}{2},-\frac{3}{2})}8_{(1,\frac{1}{2},\frac{1}{2})},\\\nonumber
15^{(2)}_{(\frac{1}{3},\frac{3}{2},\frac{3}{2})}&=&
-\frac{\sqrt{14}}{21}15_{(\frac{4}{3},1,1)}8_{(-1,\frac{1}{2},\frac{1}{2})}
+\frac{5\sqrt{7}}{21}15_{(\frac{1}{3},\frac{3}{2},\frac{3}{2})}8_{(0,1,0)}
-\frac{\sqrt{21}}{21}15_{(\frac{1}{3},\frac{3}{2},\frac{3}{2})}8_{(0,0,0)}
-\frac{5\sqrt{42}}{63}15_{(\frac{1}{3},\frac{3}{2},\frac{1}{2})}8_{(0,1,1)}\\\nonumber
&-&\frac{\sqrt{21}}{63}15_{(\frac{1}{3},\frac{1}{2},\frac{1}{2})}8_{(0,1,1)}
+\frac{4\sqrt{7}}{21}15_{(-\frac{2}{3},1,1)}8_{(1,\frac{1}{2},\frac{1}{2})},\\\nonumber
15^{(2)}_{(\frac{1}{3},\frac{3}{2},\frac{1}{2})}&=&
-\frac{\sqrt{42}}{63}15_{(\frac{4}{3},1,1)}8_{(-1,\frac{1}{2},-\frac{1}{2})}
-\frac{2\sqrt{21}}{63}15_{(\frac{4}{3},1,0)}8_{(-1,\frac{1}{2},\frac{1}{2})}
+\frac{5\sqrt{42}}{63}15_{(\frac{1}{3},\frac{3}{2},\frac{3}{2})}8_{(0,1,-1)}
+\frac{5\sqrt{7}}{63}15_{(\frac{1}{3},\frac{3}{2},\frac{1}{2})}8_{(0,1,0)}\\\nonumber
&-&\frac{\sqrt{21}}{21}15_{(\frac{1}{3},\frac{3}{2},\frac{1}{2})}8_{(0,0,0)}
-\frac{\sqrt{14}}{63}15_{(\frac{1}{3},\frac{1}{2},\frac{1}{2})}8_{(0,1,0)}
-\frac{10\sqrt{14}}{63}15_{(\frac{1}{3},\frac{3}{2},-\frac{1}{2})}8_{(0,1,1)}
-\frac{\sqrt{7}}{63}15_{(\frac{1}{3},\frac{1}{2},-\frac{1}{2})}8_{(0,1,1)}\\\nonumber
&+&\frac{4\sqrt{21}}{63}15_{(-\frac{2}{3},1,1)}8_{(1,\frac{1}{2},-\frac{1}{2})}
+\frac{4\sqrt{42}}{63}15_{(-\frac{2}{3},1,0)}8_{(1,\frac{1}{2},\frac{1}{2})},\\\nonumber
\end{eqnarray}
\begin{eqnarray}\nonumber
15^{(2)}_{(\frac{1}{3},\frac{3}{2},-\frac{1}{2})}&=&
-\frac{2\sqrt{21}}{63}15_{(\frac{4}{3},1,0)}8_{(-1,\frac{1}{2},-\frac{1}{2})}
+\frac{10\sqrt{14}}{63}15_{(\frac{1}{3},\frac{3}{2},\frac{1}{2})}8_{(0,1,-1)}
-\frac{\sqrt{7}}{63}15_{(\frac{1}{3},\frac{1}{2},\frac{1}{2})}8_{(0,1,-1)}
-\frac{\sqrt{42}}{63}15_{(\frac{4}{3},1,-1)}8_{(-1,\frac{1}{2},\frac{1}{2})}\\\nonumber
&-&\frac{5\sqrt{7}}{63}15_{(\frac{1}{3},\frac{3}{2},-\frac{1}{2})}8_{(0,1,0)}
-\frac{\sqrt{21}}{21}15_{(\frac{1}{3},\frac{3}{2},-\frac{1}{2})}8_{(0,0,0)}
-\frac{2\sqrt{14}}{63}15_{(\frac{1}{3},\frac{1}{2},-\frac{1}{2})}8_{(0,1,0)}
-\frac{5\sqrt{42}}{63}15_{(\frac{1}{3},\frac{3}{2},-\frac{3}{2})}8_{(0,1,1)}\\\nonumber
&+&\frac{4\sqrt{42}}{63}15_{(-\frac{2}{3},1,0)}8_{(1,\frac{1}{2},-\frac{1}{2})}
+\frac{4\sqrt{21}}{63}15_{(-\frac{2}{3},1,-1)}8_{(1,\frac{1}{2},\frac{1}{2})},\\\nonumber
15^{(2)}_{(\frac{1}{3},\frac{3}{2},-\frac{3}{2})}&=&
-\frac{\sqrt{14}}{21}15_{(\frac{4}{3},1,-1)}8_{(-1,\frac{1}{2},-\frac{1}{2})}
+\frac{5\sqrt{42}}{63}15_{(\frac{1}{3},\frac{3}{2},-\frac{1}{2})}8_{(0,1,-1)}
-\frac{\sqrt{21}}{63}15_{(\frac{1}{3},\frac{1}{2},-\frac{1}{2})}8_{(0,1,-1)}
-\frac{5\sqrt{7}}{21}15_{(\frac{1}{3},\frac{3}{2},-\frac{3}{2})}8_{(0,1,0)}\\\nonumber
&-&\frac{\sqrt{21}}{21}15_{(\frac{1}{3},\frac{3}{2},-\frac{3}{2})}8_{(0,0,0)}
+\frac{4\sqrt{7}}{21}15_{(-\frac{2}{3},1,-1)}8_{(1,\frac{1}{2},-\frac{1}{2})},\\\nonumber
15^{(2)}_{(\frac{1}{3},\frac{1}{2},\frac{1}{2})}&=&
\frac{\sqrt{21}}{9}15_{(\frac{4}{3},1,1)}8_{(-1,\frac{1}{2},-\frac{1}{2})}
-\frac{\sqrt{42}}{18}15_{(\frac{4}{3},1,0)}8_{(-1,\frac{1}{2},\frac{1}{2})}
+\frac{\sqrt{21}}{63}15_{(\frac{1}{3},\frac{3}{2},\frac{3}{2})}8_{(0,1,-1)}
-\frac{\sqrt{14}}{63}15_{(\frac{1}{3},\frac{3}{2},\frac{1}{2})}8_{(0,1,0)}\\\nonumber
&+&\frac{25\sqrt{7}}{252}15_{(\frac{1}{3},\frac{1}{2},\frac{1}{2})}8_{(0,1,0)}
-\frac{\sqrt{21}}{84}15_{(\frac{1}{3},\frac{1}{2},\frac{1}{2})}8_{(0,0,0)}
+\frac{\sqrt{7}}{63}15_{(\frac{1}{3},\frac{3}{2},-\frac{1}{2})}8_{(0,1,1)}
-\frac{25\sqrt{14}}{252}15_{(\frac{1}{3},\frac{1}{2},-\frac{1}{2})}8_{(0,1,1)}\\\nonumber
&+&\frac{5\sqrt{42}}{126}15_{(-\frac{2}{3},1,1)}8_{(1,\frac{1}{2},-\frac{1}{2})}
-\frac{5\sqrt{21}}{126}15_{(-\frac{2}{3},1,0)}8_{(1,\frac{1}{2},\frac{1}{2})}
+\frac{\sqrt{42}}{12}15_{(-\frac{2}{3},0,0)}8_{(1,\frac{1}{2},\frac{1}{2})},\\\nonumber
15^{(2)}_{(\frac{1}{3},\frac{1}{2},-\frac{1}{2})}&=&
\frac{7\sqrt{42}}{126}15_{(\frac{4}{3},1,0)}8_{(-1,\frac{1}{2},-\frac{1}{2})}
+\frac{\sqrt{7}}{63}15_{(\frac{1}{3},\frac{3}{2},\frac{1}{2})}8_{(0,1,-1)}
+\frac{25\sqrt{14}}{252}15_{(\frac{1}{3},\frac{1}{2},\frac{1}{2})}8_{(0,1,-1)}
-\frac{\sqrt{21}}{9}15_{(\frac{4}{3},1,-1)}8_{(-1,\frac{1}{2},\frac{1}{2})}\\\nonumber
&-&\frac{\sqrt{14}}{63}15_{(\frac{1}{3},\frac{3}{2},-\frac{1}{2})}8_{(0,1,0)}
-\frac{25\sqrt{7}}{252}15_{(\frac{1}{3},\frac{1}{2},-\frac{1}{2})}8_{(0,1,0)}
-\frac{\sqrt{21}}{84}15_{(\frac{1}{3},\frac{1}{2},-\frac{1}{2})}8_{(0,0,0)}
+\frac{\sqrt{21}}{63}15_{(\frac{1}{3},\frac{3}{2},-\frac{3}{2})}8_{(0,1,1)}\\\nonumber
&+&\frac{5\sqrt{21}}{126}15_{(-\frac{2}{3},1,0)}8_{(1,\frac{1}{2},-\frac{1}{2})}
+\frac{\sqrt{42}}{12}15_{(-\frac{2}{3},0,0)}8_{(1,\frac{1}{2},-\frac{1}{2})}
-\frac{5\sqrt{42}}{126}15_{(-\frac{2}{3},1,-1)}8_{(1,\frac{1}{2},\frac{1}{2})},\\\nonumber
15^{(2)}_{(-\frac{2}{3},1,1)}&=&
\frac{4\sqrt{7}}{21}15_{(\frac{1}{3},\frac{3}{2},\frac{3}{2})}8_{(1,\frac{1}{2},-\frac{1}{2})}
-\frac{4\sqrt{21}}{63}15_{(\frac{1}{3},\frac{3}{2},\frac{1}{2})}8_{(1,\frac{1}{2},\frac{1}{2})}
-\frac{5\sqrt{42}}{126}15_{(\frac{1}{3},\frac{1}{2},\frac{1}{2})}8_{(1,\frac{1}{2},\frac{1}{2})}
+\frac{\sqrt{7}}{7}15_{(-\frac{2}{3},1,1)}8_{(0,1,0)}\\\nonumber
&+&\frac{\sqrt{21}}{21}15_{(-\frac{2}{3},1,1)}8_{(0,0,0)}
-\frac{\sqrt{7}}{7}15_{(-\frac{2}{3},1,0)}8_{(0,1,1)}
-\frac{\sqrt{14}}{42}15_{(-\frac{2}{3},0,0)}8_{(0,1,1)}
+\frac{4\sqrt{7}}{21}15_{(-\frac{5}{3},\frac{1}{2},\frac{1}{2})}8_{(1,\frac{1}{2},\frac{1}{2})},\\\nonumber
15^{(2)}_{(-\frac{2}{3},1,0)}&=&
\frac{4\sqrt{42}}{63}15_{(\frac{1}{3},\frac{3}{2},\frac{1}{2})}8_{(-1,\frac{1}{2},-\frac{1}{2})}
-\frac{5\sqrt{21}}{126}15_{(\frac{1}{3},\frac{1}{2},\frac{1}{2})}8_{(-1,\frac{1}{2},-\frac{1}{2})}
-\frac{4\sqrt{42}}{63}15_{(\frac{1}{3},\frac{3}{2},-\frac{1}{2})}8_{(-1,\frac{1}{2},\frac{1}{2})}
-\frac{5\sqrt{21}}{126}15_{(\frac{1}{3},\frac{1}{2},-\frac{1}{2})}8_{(-1,\frac{1}{2},\frac{1}{2})}
\\\nonumber
&+&\frac{\sqrt{7}}{7}15_{(-\frac{2}{3},1,1)}8_{(0,1,-1)}
+\frac{\sqrt{21}}{21}15_{(\frac{1}{3},\frac{3}{2},\frac{1}{2})}8_{(0,0,0)}
-\frac{\sqrt{14}}{42}15_{(\frac{1}{3},\frac{1}{2},\frac{1}{2})}8_{(0,1,0)}
+\frac{2\sqrt{14}}{21}15_{(-\frac{5}{3},\frac{1}{2},\frac{1}{2})}8_{(1,\frac{1}{2},-\frac{1}{2})}\\\nonumber
&-&\frac{\sqrt{7}}{7}15_{(-\frac{2}{3},1,-1)}8_{(0,1,1)}
+\frac{2\sqrt{14}}{21}15_{(-\frac{5}{3},\frac{1}{2},-\frac{1}{2})}8_{(1,\frac{1}{2},\frac{1}{2})},\\\nonumber
15^{(2)}_{(-\frac{2}{3},1,-1)}&=&
\frac{4\sqrt{21}}{63}15_{(\frac{1}{3},\frac{3}{2},-\frac{1}{2})}8_{(-1,\frac{1}{2},-\frac{1}{2})}
-\frac{5\sqrt{42}}{126}15_{(\frac{1}{3},\frac{1}{2},-\frac{1}{2})}8_{(-1,\frac{1}{2},-\frac{1}{2})}
+\frac{\sqrt{7}}{7}15_{(-\frac{2}{3},1,0)}8_{(0,1,-1)}
-\frac{\sqrt{14}}{42}15_{(-\frac{2}{3},0,0)}8_{(0,1,-1)}\\\nonumber
&-&\frac{4\sqrt{7}}{21}15_{(\frac{1}{3},\frac{3}{2},\frac{3}{2})}8_{(-1,\frac{1}{2},\frac{1}{2})}
-\frac{\sqrt{7}}{7}15_{(-\frac{2}{3},1,-1)}8_{(0,1,0)}
+\frac{\sqrt{21}}{21}15_{(-\frac{2}{3},1,-1)}8_{(0,0,0)}
+\frac{4\sqrt{7}}{21}15_{(-\frac{5}{3},\frac{1}{2},-\frac{1}{2})}8_{(1,\frac{1}{2},-\frac{1}{2})},\\\nonumber
15^{(2)}_{(-\frac{2}{3},0,0)}&=&
\frac{\sqrt{42}}{12}15_{(\frac{1}{3},\frac{1}{2},\frac{1}{2})}8_{(-1,\frac{1}{2},-\frac{1}{2})}
-\frac{\sqrt{42}}{12}15_{(\frac{1}{3},\frac{1}{2},-\frac{1}{2})}8_{(-1,\frac{1}{2},\frac{1}{2})}
+\frac{\sqrt{14}}{42}15_{(-\frac{2}{3},1,1)}8_{(0,1,-1)}
-\frac{\sqrt{14}}{42}15_{(-\frac{2}{3},1,0)}8_{(0,1,0)}\\\nonumber
&+&\frac{\sqrt{21}}{14}15_{(-\frac{2}{3},0,0)}8_{(0,0,0)}
+\frac{\sqrt{7}}{7}15_{(-\frac{5}{3},\frac{1}{2},\frac{1}{2})}8_{(1,\frac{1}{2},-\frac{1}{2})}
+\frac{\sqrt{14}}{42}15_{(-\frac{2}{3},1,-1)}8_{(0,1,1)}
-\frac{\sqrt{7}}{7}15_{(-\frac{5}{3},\frac{1}{2},-\frac{1}{2})}8_{(1,\frac{1}{2},\frac{1}{2})},\\\nonumber
15^{(2)}_{(-\frac{5}{3},\frac{1}{2},\frac{1}{2})}&=&
\frac{4\sqrt{7}}{21}15_{(-\frac{2}{3},1,1)}8_{(-1,\frac{1}{2},-\frac{1}{2})}
-\frac{2\sqrt{14}}{21}15_{(-\frac{2}{3},1,0)}8_{(-1,\frac{1}{2},\frac{1}{2})}
-\frac{\sqrt{7}}{7}15_{(-\frac{2}{3},0,0)}8_{(-1,\frac{1}{2},\frac{1}{2})}
+\frac{\sqrt{7}}{21}15_{(-\frac{5}{3},\frac{1}{2},\frac{1}{2})}8_{(0,1,0)}\\\nonumber
&+&\frac{\sqrt{21}}{7}15_{(-\frac{5}{3},\frac{1}{2},\frac{1}{2})}8_{(0,0,0)}
-\frac{\sqrt{14}}{21}15_{(-\frac{5}{3},\frac{1}{2},-\frac{1}{2})}8_{(0,1,1)},\\\nonumber
15^{(2)}_{(-\frac{5}{3},\frac{1}{2},-\frac{1}{2})}&=&
\frac{2\sqrt{14}}{21}15_{(-\frac{2}{3},1,0)}8_{(-1,\frac{1}{2},-\frac{1}{2})}
-\frac{\sqrt{7}}{7}15_{(-\frac{2}{3},0,0)}8_{(-1,\frac{1}{2},-\frac{1}{2})}
-\frac{4\sqrt{7}}{21}15_{(-\frac{2}{3},1,-1)}8_{(-1,\frac{1}{2},\frac{1}{2})}
+\frac{\sqrt{14}}{21}15_{(-\frac{5}{3},\frac{1}{2},\frac{1}{2})}8_{(0,1,-1)}\\\nonumber
&-&\frac{\sqrt{7}}{21}15_{(-\frac{5}{3},\frac{1}{2},-\frac{1}{2})}8_{(0,1,0)}
+\frac{\sqrt{21}}{7}15_{(-\frac{5}{3},\frac{1}{2},-\frac{1}{2})}8_{(0,0,0)}.\\\nonumber
\end{eqnarray}\end{widetext}


\begin{thebibliography}{99}
\bibitem{Skyr}
T.H.R.~Skyrme, \Journal{\PRSA} {260}{127}{1961}.
\bibitem{Guad}
E.~Guadagnini, \Journal{\NPB} {236}{35}{1984}.
\bibitem{Ca_Kl}
C.~Callan and I.~Klebanov, \Journal{\NPB} {262}{365}{1985};\par
C.~Callan, K.~Hornbostel and I.~Klebanov,
\Journal{\PLB}{202}{269}{1988}.
\bibitem{rrs}
M.~Rho, D.~O.~Riska,
N.~N.~Scoccola,\Journal{\PLB}{251}{597}{1990}; D.~O. Riska, and
N.~N.~Scoccola, \Journal{\PLB}{265}{188}{1991}; M.~Rho,
D.~O.~Riska, N.~N.~Scoccola, \Journal{Z.Phys.A}{341}{343}{1992}.
\bibitem{Wall}
H.~Walliser, \Journal{\NPA} {548}{649}{1992}.
\bibitem{JAM}
E.~Jenkins, A.~V.~Manohar and M.~B.~Wise,
\Journal{\NPB}{396}{27}{1993}.
\bibitem{GLM}
Z.~Guralnik, M.~Luke and A.~V.~Manohar,
\Journal{\NPB}{390}{474}{1993}.
\bibitem{GMSS}
K.~S. Gupta, M.~Arshad~Momen, J.~Schechter and A.~Subbaraman,
\Journal{\PRD}{47}{R4835}{1993}.
\bibitem{OPM}
Y.~Oh, B.-Y.~Park and D.-P.~Min, \Journal{\PRD}{50}{3350}{1994}.
\bibitem{mss}
A.~Momen, J.~Schechter, and A.~Subbaraman,
\Journal{\PRD}{49}{5970}{1994}.
\bibitem{RGG}
A.~De~R$\acute{\mbox{u}}$jula, H.~Georgi, and S.~L.~Glashow,
\Journal{\PRD}{12}{147}{1975}.
\bibitem{RS}
D.~O.~Riska and N.~N.~Scoccola, \Journal{\PLB}{299}{338}{1993}.
\bibitem{OPM1}
Y.~Oh, B.-Y.~Park and D.-P.~Min, \Journal{\PLB}{331}{362}{1994}.
\bibitem{wise}
M.~B.~Wise, \Journal{\PRD}{45}{R2188}{1991}.
\bibitem{Chem}
M.~Chemtob, \Journal{\NPB}{256}{600}{1985}.
\end{thebibliography}
\end{document}